\pdfoutput=1
\documentclass[journal,twoside,web]{ieeecolor}
\usepackage{generic}
\usepackage{amsmath,amssymb,amsfonts}
\usepackage{graphicx}

\usepackage{subfigure}
\usepackage[lined,ruled,vlined]{algorithm2e}
\definecolor{mycommentcolor}{RGB}{191, 191, 191}

\SetCommentSty{mycomment}
\SetKwComment{Comment}{/* }{ */}
\usepackage[colorlinks=true,
    linkcolor=blue,
urlcolor=blue,citecolor=blue]{hyperref}
\usepackage{longtable}
\newlength\savewidth\newcommand\shline{\noalign{\global\savewidth\arrayrulewidth
  \global\arrayrulewidth 1.5pt}\hline\noalign{\global\arrayrulewidth\savewidth}}
\usepackage{booktabs}
\usepackage{bbding}
\usepackage{latexsym}
\usepackage{framed,multirow}
\usepackage{threeparttable}
\usepackage{xspace}

\usepackage{algpseudocode}
\usepackage{bm} 
\usepackage{bbm}

\newcommand{\ie}{\textit{i}.\textit{e}.}
\newcommand{\eg}{\textit{e}.\textit{g}.}
\newcommand{\etal}{\emph{et al.}\xspace}

\newcommand{\T}{^{\textrm T}} 

\newcommand{\eat}[1]{}

\newcommand{\tabref}[1]{Table~\ref{#1}}
\newcommand{\figref}[1]{Fig.~\ref{#1}}

\newcommand{\equref}[1]{Eq.~\eqref{#1}}
\newcommand{\secref}[1]{Sec.~\ref{#1}}

\newcommand{\tabincell}[2]{\begin{tabular}{@{}#1@{}}#2\end{tabular}}

\newcommand{\loss}[1]{\mathcal{L}_\text{#1}}
\newcommand{\expe}[1]{\mathbb{E}_{#1}}  
\newcommand{\lamda}[1]{\lambda_\text{#1}}   

\newcommand{\best}[1]{\textbf{#1}}

\newcommand{\Map}[1]{\mathcal{#1}}
\newcommand{\Set}[1]{\mathcal{#1}}
\newcommand{\vct}[1]{\boldsymbol{#1}} 
\newcommand{\mat}[1]{\boldsymbol{#1}} 
\newcommand{\img}[1]{\boldsymbol{#1}} 
\DeclareMathOperator*{\argmin}{arg\,min}

\newcommand{\methodname}{{HOPE}\xspace}

\usepackage{tikz}
\usepackage{textcomp}
\usepackage{lipsum}
\newcommand\copyrighttext{%
  \footnotesize \textcopyright 2024 IEEE. Personal use of this material is permitted.
  Permission from IEEE must be obtained for all other uses, in any current or future
  media, including reprinting/republishing this material for advertising or promotional
  purposes, creating new collective works, for resale or redistribution to servers or
  lists, or reuse of any copyrighted component of this work in other works.
  }
\newcommand\copyrightnotice{%
\begin{tikzpicture}[remember picture,overlay]
\node[anchor=south,yshift=3pt] at (current page.south) {\fbox{\parbox{\dimexpr\textwidth-\fboxsep-\fboxrule\relax}{\copyrighttext}}};
\end{tikzpicture}%
}

\def\BibTeX{{\rm B\kern-.05em{\sc i\kern-.025em b}\kern-.08em
    T\kern-.1667em\lower.7ex\hbox{E}\kern-.125emX}}
\begin{document}
\title{HOPE: Hybrid-granularity Ordinal Prototype Learning for Progression Prediction of \\Mild Cognitive Impairment}

\author{
Chenhui Wang, 
Yiming Lei,~\IEEEmembership{Member, IEEE}, 
Tao Chen, 
Junping Zhang,~\IEEEmembership{Senior Member, IEEE},\\
Yuxin Li,
Hongming Shan,~\IEEEmembership{Senior Member, IEEE}
and the Alzheimer’s Disease Neuroimaging Initiative
}


\maketitle

\copyrightnotice

\begin{abstract}
Mild cognitive impairment (MCI) is often at high risk of progression to Alzheimer’s disease (AD). 
Existing works to identify the progressive MCI (pMCI) typically require MCI subtype labels, pMCI vs. stable MCI (sMCI), determined by whether or not an MCI patient will progress to AD after a long follow-up. 
However, prospectively acquiring MCI subtype data is time-consuming and resource-intensive; the resultant small datasets could lead to severe overfitting and difficulty in extracting discriminative information.
Inspired by that various longitudinal biomarkers and cognitive measurements present an ordinal pathway on AD progression, we propose a novel Hybrid-granularity Ordinal PrototypE learning (\methodname) method to characterize AD ordinal progression for MCI progression prediction.
First, \methodname learns an ordinal metric space that enables progression prediction by prototype comparison.
Second, \methodname leverages a novel hybrid-granularity ordinal loss to learn the ordinal nature of AD via effectively integrating instance-to-instance ordinality, instance-to-class compactness, and class-to-class separation.
Third, to make the prototype learning more stable, \methodname employs an exponential moving average strategy to learn the global prototypes of NC and AD dynamically.
Experimental results on the internal ADNI and the external NACC datasets demonstrate the superiority of the proposed \methodname over existing state-of-the-art methods as well as its interpretability.
Source code is made available at \url{https://github.com/thibault-wch/HOPE-for-mild-cognitive-impairment}.

\end{abstract}
\section{Introduction}
As the most common cause of dementia, Alzheimer's disease (AD) presents irreversible and progressive characteristics whilst resulting in severe cognitive impairment and behavior issues~\cite{winblad2016defeating}. Mild cognitive impairment (MCI), the prodromal stage of AD, is receiving increasing attention due to the high risks of progression to AD. Depending on the DSM-V criteria~\cite{american2013diagnostic} of whether or not an MCI patient will progress to AD after a period of follow-up (typically 36 months),  the current MCI patients can be classified as two finer-grained subtypes:  progressive MCI (pMCI) and stable MCI (sMCI). Unlike the classification between AD and Normal Cognition (NC)~\cite{qiu2020development} that can be easily achieved by neuroimaging differences among individuals, distinguishing pMCI from sMCI poses a greater challenge due to their high degree of intra-similarity~\cite{huang2019diagnosis}. Therefore, identifying the pMCI as early as possible is critical for subsequent intervention.

\begin{figure}[!t]
	\centering  
	\includegraphics[width=1\linewidth]{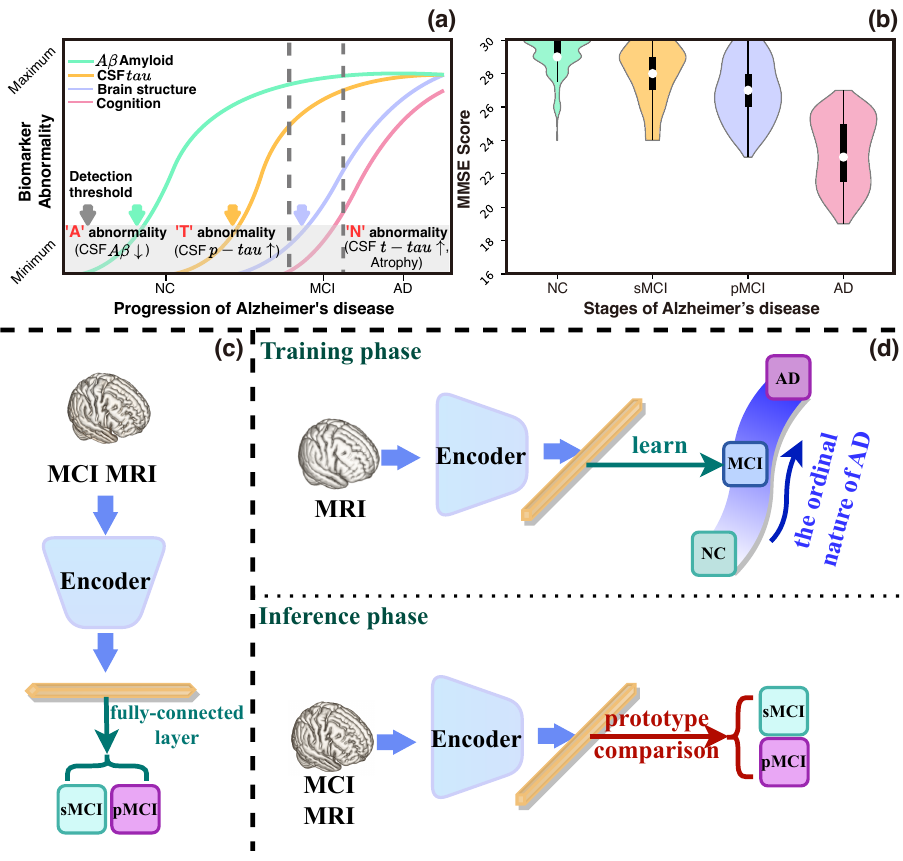}  
	\caption{Illustration of the ordinal development of AD and comparison between existing works and our \methodname for MCI progression prediction. (\textbf{a}) Abnormal development order of different biomarkers. (\textbf{b}) Progressive cognitive decline illustrated in the violin plot of Mini-Mental State Examination (MMSE) scores extracted from the internal ADNI dataset. (\textbf{c}) Existing works require MCI subtype labels, \ie~sMCI and pMCI, to predict MCI progression. (\textbf{d}) Our \methodname leverages the data across subjects with coarse-grained labels, \ie~NC$\rightarrow$MCI$\rightarrow$AD, to learn the ordinal nature of AD and predict MCI progression by prototype comparison.
 }
 \label{fig:main_sum}
\end{figure}

Existing MCI progression prediction works~\cite{pan2020early,beheshti2017classification,ashtari2022multi} typically require MCI subtype labels and differentiate them based on the progressive rate, as shown in \figref{fig:main_sum}(c). However, prospectively acquiring MCI subtype data is time-consuming and resource-intensive, which leads to relatively small labeled datasets, resulting in amplified overfitting and challenges in extracting discriminative information for real-world scenarios. Recently, some previous studies revealed that the models trained by NC and AD data perform better than those trained by sMCI and pMCI data for predicting the progression of MCI~\cite{huang2019diagnosis,falahati2014multivariate,kwak2021subtyping,kwak2022identifying}. However, these studies only considered that differences between sMCI and pMCI are analogical to those between NC and AD, ignoring the global ordinal nature of AD progression.

During the development of AD, various longitudinal biomarkers and cognitive measurements present a unidirectional ordinal pathway~\cite{guo2021longitudinal,mill2022recent} based on the $A\beta$, $tau$, and neurodegeneration (A/T/N) biomarker classification system~\cite{jack2018nia,zhang2022influence,tian2023international}, as shown in \figref{fig:main_sum}(a-b). Aging inevitably leads to brain atrophy, regardless of AD. Meanwhile, the progression rate of brain atrophy gradually increases as the disease progresses~\cite{vos2015prevalence,parnetti2019prevalence}. This phenomenon explains why, for a subject's longitudinal MRIs, the later image has higher prediction accuracy than earlier ones. As an extreme example, if an NC individual maintains long-term stability, it is primarily due to a very slow brain atrophy rate.

Based on the ordinal disease development of AD, we regard the extensive cross-sectional data collected from subjects with different coarse-grained diagnosis labels, ranging from NC to MCI to AD, as the ``latent'' longitudinal data specific to the entire disease duration of the AD cohort. Thus, we can leverage additional subjects of NC and AD to help predict MCI progression by learning the whole ordinal nature of AD. Towards this end, we propose a novel \textbf{H}ybrid-granularity \textbf{O}rdinal \textbf{P}rototyp\textbf{E} learning (\methodname) method to perform MCI fine-grained label (sMCI and pMCI) classification with only coarse-grained label data (NC, MCI, and AD) through learning the ordinal nature of AD, yielding a more applicable, generalizable algorithm towards real-world scenarios, as shown in \figref{fig:main_sum}(d). More specifically, we propose a novel hybrid-granularity ordinal loss based on the prior knowledge of AD progression order. This loss function is meticulously crafted to directly rank subjects in the feature space, achieving instance-to-instance ordinality while maintaining instance-to-class compactness and class-to-class separation. Moreover, we propose an online EMA prototype learning strategy to dynamically update the prototype during training, enhancing the stability of prototype learning and achieving higher performance.

The contributions of this work are summarized as follows.
\begin{enumerate}
    \item We propose a novel hybrid-granularity ordinal prototype learning method to learn an ordinal metric space that enables MCI progression prediction by prototype comparison. To the best of our knowledge, this is the \emph{first} study focusing on the ordinal progression of AD in deep learning-based methods.
    \item We propose a novel hybrid-granularity ordinal loss to learn the global ordinal nature of AD via effectively integrating instance-to-instance ordinality, instance-to-class compactness, and class-to-class separation.
    \item We employ the Exponential Moving Average (EMA) strategy to dynamically learn the global prototypes of NC and AD in an online manner to make the prototype learning more stable.
    \item Experimental results on the internal ADNI and external NACC datasets show that our \methodname outperforms the recently published state-of-the-art methods and has better generalizability, and that the regions of interest identified by our \methodname are associated with the worsening of AD.
\end{enumerate}
The remainder of this paper is organized as follows. We briefly review related works in \secref{sec:related}.
We then detail the proposed methodology in \secref{sec:method}. We elaborate on the experimental setup and results in \secref{sec:results}. Finally, \secref{sec:dicussion} presents the discussion of our \methodname, followed by a conclusion in \secref{sec:conclusion}.
\section{Related Work}

In this section, we briefly review related works on the progression prediction of MCI, fine-grained classification, and ordinal estimation.
\label{sec:related}
\subsection{Progression Prediction of MCI}

Most of the studies on AD focus mainly on differentiating AD from NC~\cite{qiu2020development,chen2019med3d,zhang2021explainable,bien2018deep,wang2024joint}. However, there is relatively little work on predicting the progression of MCI due to their high degree of intra-similarity and the high cost associated with prospectively acquiring longitudinal data. 
For example, Ashtari~\etal proposed a multi-stream convolutional neural network (CNN) to classify sMCI and pMCI~\cite{ashtari2022multi}. 
Existing MCI progression prediction works~\cite{beheshti2017classification,ashtari2022multi, zichao2023} consider that the criterion of differentiating sMCI and pMCI should be based on the progressive rate and try to mine discriminative information \emph{solely} from these two MCI subtype groups.

Recently, some studies have shown that models trained by NC and AD outperform those trained by sMCI and pMCI for predicting the progression of MCI~\cite{huang2019diagnosis,falahati2014multivariate,ashtari2022multi,kwak2021subtyping,kwak2022identifying,gao2023hybrid}. Huang~\etal proposed a novel CNN to fuse the multi-modality information from T1-MRI and FDG-PET images around the hippocampus area for predicting MCI progression~\cite{huang2019diagnosis}.  Their method is trained using data from NC and AD and then tested in the sMCI and pMCI. Similarly, Kwak~\etal leveraged a 3D DenseNet model to differentiate NC and AD subjects based on whole brain GM density maps and is subsequently tested for classifying sMCI and pMCI~\cite{kwak2022identifying}.  However, these studies only considered that the differences between sMCI and pMCI are analogical to those between NC and AD, ignoring the whole ordinal nature of AD progression. Unlike them, our \methodname distinguishes pMCI from sMCI by learning the progression rate from the data across subjects with different coarse-grained labels, \ie~NC$\rightarrow$MCI$\rightarrow$AD.

\subsection{Fine-grained Classification}
Due to the small inter-class variations between fine-grained categories and the large intra-class variations in individual differences~\cite{wei2021fine}, numerous researchers have paid their attention on the fine-grained classification field~\cite{yang2018learning,nauta2022neural,shu2022improving,kim2022vit}. However, existing state-of-the-art (SOTA) fine-grained classification methods 
mainly focus on fine-grained labeled data to mine discriminative information, which can be roughly divided into two general categories~\cite{wei2021fine}:  recognition by localization-classification subnetworks~\cite{yang2018learning,nauta2022neural,shu2022improving,kim2022vit}, and recognition by end-to-end feature encoding~\cite{song2022eigenvalues}. 

In the context of recognition by localization-classification subnetworks, Yang~\etal proposed a novel self-supervision mechanism to effectively localize informative regions without the need for bounding-box/part annotations to find fine-grained discriminative features~\cite{yang2018learning}. Concerning recognition by end-to-end feature encoding, Song~\etal proposed a network branch dedicated to amplifying the significance of small eigenvalues using the global covariance pooling layer, effectively capturing inter-class differences~\cite{song2022eigenvalues}. Unlike these fine-grained classification methods that mine the fine-grained discriminative information among subcategories (sMCI and pMCI), our \methodname takes the ordinal nature of AD into account and uses larger inter-class variations among coarse-grained labels (NC, MCI, and AD) to distinguish them. 

\subsection{Ordinal Estimation}
With the development of deep learning (DL), there are already lots of DL-based methods to mine the ordinal nature of specific problems, such as age estimation~\cite{niu2016ordinal,zhu2021convolutional,lei2023core,wang2023controlling} and aesthetic quality control classification~\cite{rosati2022novel}. Niu~\etal proposed an ordinal regression CNN to transform the age estimation problem into a series of binary classification sub-problems to make the predicted label space ordinal~\cite{niu2016ordinal}.
Wang~\etal proposed a novel Constrained Proxies Learning (CPL) method to learn the proxy for each ordinal age class and then adjusts the global layout of classes by making these proxies to satisfy specific constraint conditions (\eg~Poisson or Bernoulli distribution)~\cite{wang2023controlling}. Recently, considerable attention has been given to exploiting the ordinal nature of specific diseases~\cite{lei2022meta,lei2023clip}. Lei~\etal proposed a novel meta ordinal regression forest (MORF) method for lung nodule malignancy prediction by combining CNN and differential forest in a meta-learning framework~\cite{lei2022meta}.

Although statistical models, such as event-based model (EBM)~\cite{fonteijn2012event}, were already established in 2012 for AD diagnosis based on the AD ordinal nature. However, the DL-based ordinal method on AD remains unexplored. Moreover, the proposed hybrid-granularity ordinal loss is \emph{directly} applied to the feature space, and \emph{solely} relies on the prior knowledge of the disease progression order to rank different subjects, which can effectively learn the ordinal nature of AD.

\section{Methodology}
\label{sec:method}
In this section, we first describe the problem definition and overview of the proposed \methodname in Secs.~\ref{sec:problem} and~\ref{sec:overview}, respectively. Then, we detail the proposed hybrid-granularity ordinal loss in \secref{sec:loss_ordinal}. Moreover, we present our online EMA prototype update strategy in \secref{sec:ema}. Finally, we present the objective function in \secref{sec:object}.

\subsection{Problem Definition}
\label{sec:problem}

Suppose that we have a dataset $\Map{D}=\{\Map{X},\Map{Y}\}$ with $N$ input 3D T1 weighted MRI scans $\Map{X}=\{\img{x}_i\}_{i=1}^N$ and their corresponding ordinal coarse-grained diagnosis labels $\Map{Y}=\{y_i\}_{i=1}^N$. Here, $y_i \in\{1, \ldots, K\}$ ($K=3$); $y=1$, $2$, and $3$ represent NC, MCI, and AD, respectively.
The feature representations of $\mathcal{X}$, $\Map{Z}=\{\vct{z_i}\}_{i=1}^N$, are generated by an encoder $f_{\vct{\theta}}(\cdot)$ with the parameters $\vct{\theta}$; \ie, $\vct{z}_i=f_{\vct{\theta}}(\vct{x}_i)$. Our \methodname aims to distinguish pMCI from sMCI by learning the global ordinal relationship of AD progression from the mapping $h(\cdot):\Map{X}\rightarrow\Map{Z}\rightarrow\Map{Y}$ (from images to features to ordinal labels). Here, we use $\Set{B}=\{(\img{x}_i, y_i)\}_{i=1}^M$  to represent a mini-batch of size $M$.

\subsection{Overview of the Proposed \methodname}
\label{sec:overview}
\begin{figure*}[!t]
\centering
\label{fig:rankloss}
\centerline{\includegraphics[width=\textwidth]{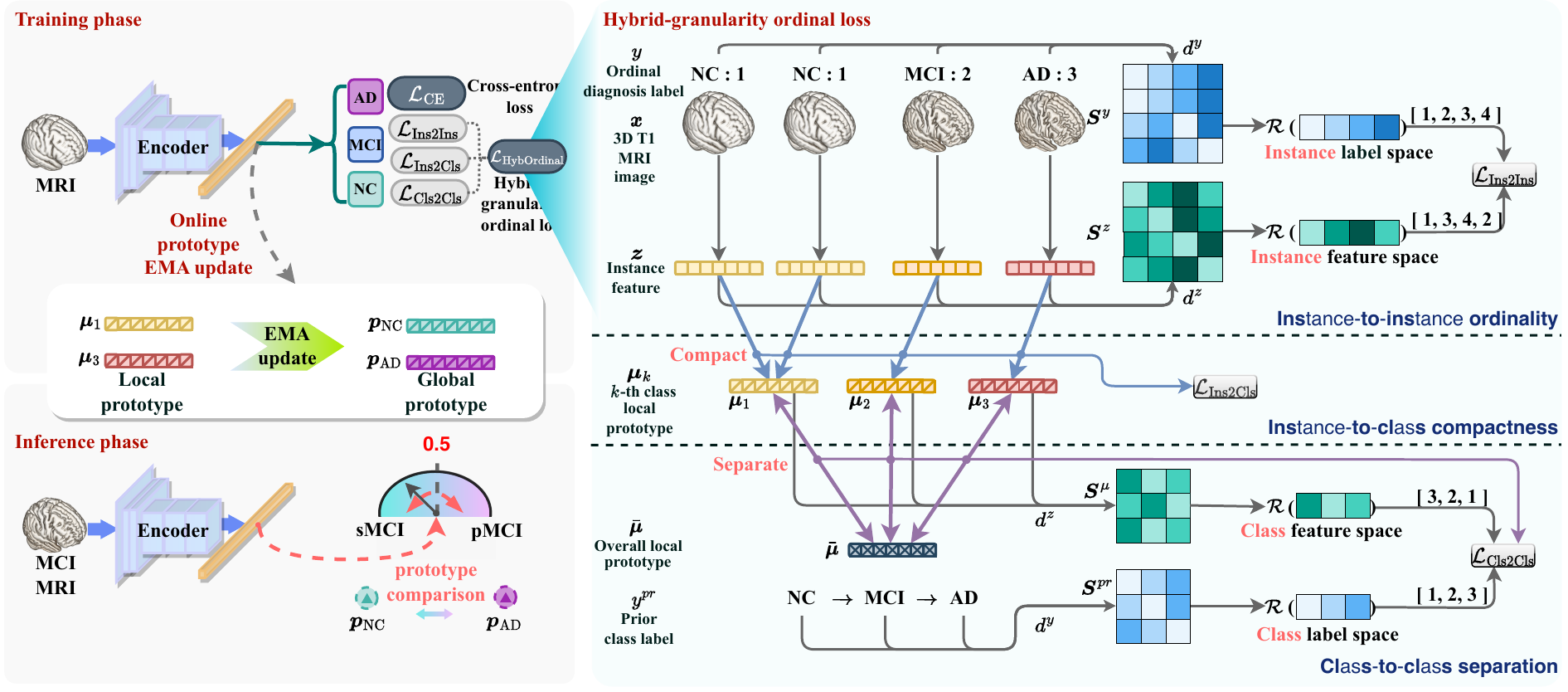}}
\caption{Overview of our hybrid-granularity ordinal prototype learning method for MCI progression prediction. Our \methodname learns the ordinal nature of AD by a novel hybrid-granularity ordinal loss and dynamically learns global prototypes by the online EMA prototype update strategy during training. The subtype label of the to-be-predicted MCI subject can be inferred by comparing it with the learned NC and AD global prototypes.}
\end{figure*}

Based on the ordinal nature of AD, we propose a novel hybrid-granularity ordinal prototype learning method to learn an ordinal metric space for predicting the progression of MCI, as shown in \figref{fig:rankloss}. 

During the training phase, our \methodname randomly samples pairs of 3D T1 weighted MRI images and labels $(\img{x}_i,y_i)$ from different classes, ranging from NC to MCI to AD, as input in a stratified manner~\cite{neyman1934two,cochran1977sampling}. Our \methodname uses cross-entropy (CE) loss to classify them through a fully-connected (FC) layer and leverages the proposed hybrid-granularity loss to learn the ordinal nature of AD via effectively integrating instance-to-instance ordinality, instance-to-class compactness, and class-to-class separation. At the same time, to make the prototype learning more stable, we employ the EMA update strategy to dynamically learn global prototypes of NC and AD. 

During the inference phase, we compare the feature of the to-be-predicted MCI subject with the learned global NC and AD prototypes to infer whether the subject will progress to AD in the future. 

In the following, we detail the proposed hybrid-granularity ordinal loss.

\subsection{Hybrid-granularity Ordinal Loss}
\label{sec:loss_ordinal}

To predict the progression of MCI by learning the inherent ordinality of AD, we propose a novel hybrid-granularity ordinal loss to rank different subjects between feature and label spaces across hybrid granularities. In the following, we detail three different granular loss components: instance-to-instance ordinality, instance-to-class compactness, and class-to-class separation.

\subsubsection{Instance-to-instance ordinality}
For instances across different AD stages, our \methodname uses the proposed instance-to-instance loss to align the ordinal relationship between the instance label space $\mathcal{Y}$ and instance feature space $\mathcal{Z}$  by similarity ranking~\cite{gong2022ranksim}.

Let $\mat{S}^y\in\mathbb{R}^{M\times M}$ be the pairwise instance-granularity similarity matrix in the ordinal label space across all instances in $\Set{B}$. The $(i,j)$-{th} entry in $\mat{S}^y$ measures the similarity between $y_i$ and $y_j$, defined as:
\begin{align}
    \label{equ:simi_y}
    \mat{S}^y_{i,j}=d^y(y_i,y_j),
\end{align}
where $d^y$ is a negative absolute distance. 
Analogously, a pairwise instance-granularity similarity matrix $\mat{S}^z\in\mathbb{R}^{M\times M}$ in instance feature space can be defined as:
\begin{align}
    \label{equ:simi_z}
    \mat{S}^z_{i,j}=d^z(\vct{z}_i,\vct{z}_j)=d^z(f_{\vct{\theta}}(\img{x}_i),f_{\vct{\theta}}(\img{x}_j)),
\end{align}
where $d^z$ is a cosine similarity.

The proposed instance-to-instance loss is to encourage the instance feature $\vct{z}$ to have the same rank as its instance ordinal label $y$ by minimizing the mean squared error between the ranks of instance label $\vct{y}$ and instance feature $\vct{z}$ as follows:
\begin{align}
\label{equ:ins2ins}
\loss{Ins2Ins}=\frac{1}{M}\sum_{i=1}^{M} \left\|\mathcal{R}(\mat{S}^y_{[i,:]})-\mathcal{R}(\mat{S}^z_{[i,:]})\right\|_2^2,
\end{align}
where $\mathcal{R}(\vct{a})_i=1+\sum_{j\neq i}\mathbbm{1}(\vct{a}_j>\vct{a}_i)$ is a ranking function with $\mathbbm{1}(\cdot)\in\{1,0\}$ being an indicator function, $\left\|\cdot\right\|_2$ represents the Euclidean norm, and the subscript $[i,:]$ denotes the $i$-th row of a matrix. Of note, compared to~\cite{gong2022ranksim}, for subjects within the same stages of each mini-batch, the ranking function $\mathcal{R}$ determines that the instances with lower indices are ranked first, making the learned features more robust.

In fact, \equref{equ:ins2ins} is difficult to optimize because of the non-differentiability of the piecewise constant ranking operation $\mathcal{R}$. Because the gradient of $\mathcal{R}(\vct{a})$ with respect to $\vct{a}$ is zero almost everywhere, we recast the ranking operation as the minimizer of a linear combinatorial objective~\cite{rolinek2020optimizing}:
\begin{align}
\label{equ:argcombine}
    \mathcal{R}(\vct{a})=\argmin_{\vct{\pi}\in\mathcal{P}} \vct{a}\T \vct{\pi},
\end{align}
where $\mathcal{P}$ is the set of all permutation vectors $\vct{\pi}$. To efficiently perform backpropagation through blackbox combinatorial solvers~\cite{poganvcic2020differentiation}, we construct a continuous linear interpolation to obtain an informative gradient from the piecewise constant loss landscape.

\subsubsection{Instance-to-class compactness}
The instance-to-class loss is to achieve high intra-class compactness in the feature space. Here, we use $\Set{Z}_k = \{\vct{z}_i | y_i=k\}$ to denote the feature set belonging to the $k$-th class within the mini-batch. $\vct{\mu}_k$ denotes the local prototype of $\Set{Z}_k$, defined as:
\begin{align}
\label{equ:mean value1}
    \vct{\mu}_k&=\frac{1}{\left|\Set{Z}_k\right|} \sum_{\vct{z}_i\in \Set{Z}_k} \vct{z}_i,
\end{align}
where $\left|\cdot\right|$ refers to the cardinality of the corresponding feature set. Finally, the instance-to-class loss $\loss{Ins2Cls}$ is calculated as:
\begin{align}
\label{equ:ins2cls}
\loss{Ins2Cls}=\frac{1}{d}\sum_{k=1}^{K}\sum_{\vct{z}_i\in \Set{Z}_k}\left\|\vct{z}_i-\vct{\mu}_k\right\|_2^2,
\end{align}
where $d$ denotes the dimension of the feature $\vct{z}_i$.

\subsubsection{Class-to-class separation} Furthermore, the class-to-class loss is designed to enhance the inter-class separability of the learned features and reinforce the ordinal nature based on the prior AD progression knowledge.

To enlarge the distance between overall feature representations, analogous to computing $k$-th class local prototype $\vct{\mu}_k$, we calculate the overall local prototype $\vct{\mu}$ as follows:
\begin{align}
    \label{equ:mean value2}
   \bar{\vct{\mu}}&=\frac{1}{
    \left|\Set{Z}\right|} \sum_{\vct{z}_i\in \Set{Z}}  {\vct{z}_i}.
\end{align}

At the same time, to reinforce the ordinal nature, we use the prior AD progression knowledge, \ie~NC$\rightarrow$MCI$\rightarrow$AD, to define the prior class labels $y^{pr}_i=\{1, \ldots, K\}$ ($K=3$); the values $1$, $2$, and $3$ denote NC, MCI, and AD, respectively. Then, the pairwise similarity matrices in the class label and  class feature spaces, \ie~$\mat{S}^{pr}\in\mathbb{R}^{K\times K}$ and $\mat{S}^{\mu}\in\mathbb{R}^{K\times K}$, are defined as:
\begin{align}
    \label{equ:simi_pr}
    \mat{S}^{pr}_{i,j}=d^y(y_i^{pr},y_j^{pr}), \quad
    \mat{S}^{\mu}_{i,j}=d^z(\vct{\mu}_i,\vct{\mu}_j),
\end{align}
where $\vct{\mu}_i$ and $\vct{\mu}_j$ are calculated based on \equref{equ:mean value1}. 

Finally, the class-to-class loss $\loss{Cls2Cls}$ is calculated as:
\begin{align}
    \loss{Cls2Cls}&=\frac{d}{\sum_{k=1}^{K}\left|\Set{Z}^k\right|\left\|\vct{\mu}_k-\bar{\vct{\mu}}\right\|_2^2}\nonumber\\
    &+\!\frac{1}{K}\!\sum_{i=1}^{K}\!\left\|\mathcal{R}(\mat{S}^{pr}_{[i,:]})\!-\!\mathcal{R}(\mat{S}^{\mu}_{[i,:]})\right\|_2^2\!,
    \label{equ:cls2cls}
\end{align}
where the first term is to encourage classes to be dispersed, and the second is to reinforce the learned ordinal nature.

In summary, the proposed hybrid-granularity ordinal loss is to learn the ordinal nature of AD by integrating instance-to-instance ordinality, instance-to-class compactness, and class-to-class separation, which is defined as:
\begin{align}
\label{equ:HybOrdinal}\loss{HybOrdinal}=\loss{Ins2Ins}+\loss{Ins2Cls}+\loss{Cls2Cls}.
\end{align}

\subsection{Online EMA Prototype Update Strategy}
\label{sec:ema}
Here, to make the prototype learning more stable, we employ the EMA strategy to update the global prototypes of NC and AD in an online fashion during training. First, we maintain two global prototypes,  $\vct{p}_\mathrm{NC}$ for NC and $\vct{p}_\mathrm{AD}$ for AD, which are initiated as zero vectors. With the training iteration progressing, the global prototypes are updated with  local prototypes within each mini-batch, given as follows:
\begin{align}
\begin{cases}\label{emap_2}
\vct{p}_\mathrm{NC}=\sigma\frac{\vct{p}_\mathrm{NC}}{\left\|\vct{p}_\mathrm{NC}\right\|_2}
+(1-\sigma)\frac{\vct{\mu}_1}{\left\|\vct{\mu}_1\right\|_2},\\
\vct{p}_\mathrm{AD}=\sigma\frac{\vct{p}_\mathrm{AD}}{\left\|\vct{p}_\mathrm{AD}\right\|_2}+(1-\sigma)\frac{\vct{\mu}_3}{\left\|\vct{\mu}_3\right\|_2},
\end{cases}
\end{align}
where $\sigma\in(0,1)$ is the momentum decay of EMA, and the $\vct{\mu}_1$ and $\vct{\mu}_3$ denote the local prototype of NC $(k=1)$ and AD $(k=3)$ within mini-batch, as calculated by \equref{equ:mean value1}. We note that once the training of the proposed \methodname is completed, the inference speed of the proposed \methodname is as fast as that of the conventional diagnosis network based on an FC layer without extra prototype calculating time.

\begin{algorithm}[t]
\label{alg:train}
\small
    \SetAlgoLined
    \textbf{Require:} pairs of 3D MRI images and labels ($\Map{X},\Map{Y}$)\\
    \textbf{Ensure:} optimize the encoder parameters $\vct{\theta}$\\
    \For{each mini-batch}{
        \tcp{Randomly sample data in a stratified manner}
        \textbf{Sample} $\Set{D}=\{(\img{x}_i, y_i)\}_{i=1}^M$ from $(\Map{X},\Map{Y})$\;
        \tcp{Local prototypes and similarity matrices calculation}
         $\vct{\mu}_1$, $\vct{\mu}_2$, $\vct{\mu}_3$, and $\bar{\vct{\mu}}\leftarrow$ Eqs.~(\ref{equ:mean value1}) and~(\ref{equ:mean value2})\;
         $\mat{S}^y$, $\mat{S}^z$, $\mat{S}^{pr}$, and $\mat{S}^{\mu}\leftarrow$ Eqs.~(\ref{equ:simi_y}),~(\ref{equ:simi_z}),~and (\ref{equ:simi_pr})\;
        \tcp{Hybrid-granularity ordinal loss calculation}
        \textbf{Calculate} $\loss{Ins2Ins}$ using $\mat{S}^y$ and $\mat{S}^z\leftarrow$~\equref{equ:ins2ins}\;
        \textbf{Calculate} $\loss{Ins2Cls}$ using $\vct{\mu}_1$, $\vct{\mu}_2$, and $\vct{\mu}_3\leftarrow$~\equref{equ:ins2cls}\;
       \textbf{Calculate} $\loss{Cls2Cls}$ using $\vct{\mu}_1$, $\vct{\mu}_2$, $\vct{\mu}_3$, $\bar{\vct{\mu}}$, $\mat{S}^{pr}$, and $\mat{S}^{\mu}\leftarrow$~\equref{equ:cls2cls}\;
       \textbf{Calculate} $\loss{HybOrdinal}$ using $\loss{Ins2Ins}$, $\loss{Ins2Cls}$, and $\loss{Cls2Cls}\leftarrow$ \equref{equ:HybOrdinal}\;
    \tcp{Loss landscape continuous interpolation per~\cite{poganvcic2020differentiation}}
    \textbf{Interpolate} $\triangledown_{\vct{\theta}}\loss{HybOrdinal}$ landscape\;
        \tcp{Cross-entropy loss calculation}
       \textbf{Calculate} $\loss{CE}$\;
        \tcp{Parameters update}
       \textbf{Calculate} $\loss{Total}$ using $\loss{CE}$ and $\loss{HybOrdinal}\leftarrow$ \equref{equ:loss_final}\;
        \textbf{Update} $\vct{\theta}\leftarrow\triangledown_{\vct{\theta}}\loss{Total}$\;
        \tcp{Global prototypes update}
        \textbf{Update} $\vct{p}_{\mathrm{NC}}$ and $\vct{p}_{\mathrm{AD}}$ using $\vct{\mu}_1$ and $\vct{\mu}_3\leftarrow$ Eq.~\eqref{emap_2}\;
    }
    \caption{Training procedure of \methodname.}
\end{algorithm}

\subsection{Objective Function}
\label{sec:object}
In the following, we demonstrate the objective function of the proposed \methodname. During the training phase, our \methodname randomly samples pairs of 3D T1 weighted MRI images and labels $(\img{x}_i, y_i)$ from different coarse-grained labels, ranging from NC to MCI to AD, as input in a stratified manner~\cite{neyman1934two,cochran1977sampling}. The cross-entropy (CE) loss is used to classify them: $\loss{CE}=-\expe{(y_i,y_i^*)}y_i\log(y_i^*)$, where $y_i, y_i^*$ are the ground-truth and predicted labels, respectively.

Finally, the overall objective function is defined as:
\begin{align}
\label{equ:loss_final}
\loss{Total}=\loss{CE}+\lamda{HybOrdinal}\loss{HybOrdinal},
\end{align}
where the hyperparameter $\lamda{HybOrdinal}$ is to balance the weights between CE loss and the proposed hybrid-grannularity ordinal loss. Note that the overall objective function is calculated per mini-batch. The overall training process is shown in Alg.~\ref{alg:train}.

\begin{table*}[t]
	\centering
		\caption{Demographic information of the subjects used in the experiment.}
  \label{tab:demographic_info}
		\begin{tabular*}{1\linewidth}{@{\extracolsep{\fill}}ccccccccccc}
	\shline
   \multirow{2}{*}{\textbf{Datasets}}&\multirow{2}{*}{\textbf{Subtypes}} &\textbf{Subject}&
   \multicolumn{2}{c}{\textbf{Gender}}&
   \multicolumn{2}{c}{\textbf{Age}}& 
   \multicolumn{2}{c}{\textbf{MMSE}} & 
  \multicolumn{2}{c}{\textbf{APOE4}}  \\ 
    &&no.&male&(\%)&mean&(std)&mean&(std)&positive &(\%)\\
\cline{1-3}\cline{4-5}\cline{6-7}\cline{8-9}\cline{10-11}
\multirow{4}{*}{\textbf{ADNI}}&
NC   &317   &148&(46.69)   &75.604&(6.231)   &28.588&(2.573)   &85&(26.81) \\
&sMCI   &317   &174&(54.89)   &72.387&(7.529)   &27.902&(1.724)   &138&(43.53)   \\
&pMCI   &169   &90&(53.25)   &73.547&(7.370)   &26.917&(1.807)   &117&(69.23)   \\
&AD   &247   &129&(52.23)   &74.847&(7.934)   &22.495&(3.321)   &157&(63.56) \\
\cline{1-3}\cline{4-5}\cline{6-7}\cline{8-9}\cline{10-11}
\multirow{2}{*}{\textbf{NACC}}&sMCI   &76   &43&(56.58)   &72.012&(7.737)   &27.699&(1.626)   &25&(32.89)   \\
&pMCI   &60   &34&(56.66)   &74.567&(7.674)   &27.037&(1.738)   &31&(51.67)   \\
\shline
\end{tabular*}
\end{table*}

During the inference stage, we compare the feature of the to-be-predicted MCI subject $\vct{x}^q$ with learned NC and AD global prototypes, $\vct{p}_\mathrm{NC}$ and $\vct{p}_\mathrm{AD}$,  to infer whether or not the MCI subject will progress to AD in the future. The probability of $\vct{x}^q$ progressing to AD, \ie~pMCI, is defined as:
\begin{align}
    p(y^q=\text{pMCI})= \frac{\exp \Big(d^z\left(f(\vct{x}^q), \vct{p}_\mathrm{AD}\right)\Big)}
    {\sum\limits_{\vct{p}\in\{\vct{p}_\mathrm{NC}, \vct{p}_\mathrm{AD}\}}\exp \Big(d^z\left(f(\vct{x}^q), \vct{p}\right)\Big)}.
\end{align}

Based on the ordinal development of AD, we define the differentiation threshold as 0.5. If the MCI feature is closer to the AD global prototype ($p>0.5$), it is likely to progress into AD and is therefore classified as pMCI. Similarly, if it is closer to the NC global prototype ($p\le0.5$), it is unlikely to progress into AD and is therefore classified as sMCI.

\section{Experiments}
\label{sec:results}
In this section, we first elaborate on the experimental setup in \secref{sec:setup}. Then, we compare our \methodname with other 18 SOTA competing methods on the same internal ADNI and external NACC testing sets in \secref{sec:exp_sota}. Furthermore, we present  detailed analyses of interpretable visualization results and ablation studies in Secs.~\ref{sec:interpret} and \ref{sec:ablation}, respectively. Finally, we present extension results for the case where substantial subtype labeled data are available in \secref{sec:fine-tune}.

\subsection{Experimental Setup}
\label{sec:setup}

This subsection first demonstrates datasets and then presents the preprocessing pipeline, followed by implementation details and competing methods used in the experiments.

\subsubsection{Datasets}
The data used in this experiment are obtained from the Alzheimer's Disease Neuroimaging Initiative (ADNI) and National Alzheimer's Coordinating Center (NACC) datasets. Resources and data in ADNI are from North America. ADNI researchers collect, validate, and utilize data, including MRI and PET images, genetics, cognitive tests, CSF, and blood biomarkers, as predictors of AD~\cite{petersen2010alzheimer}. Over the past two decades, NACC has been built in collaboration with more than 42 Alzheimer’s Disease Research Centers (ADRCs) in the US~\cite{beekly2004national}. All study participation protocols are reviewed by each subject’s local review committee and signed with the subject’s consent~\cite{petersen2010alzheimer}. The demographic information of the subjects used in the experiment is summarized in \tabref{tab:demographic_info}. Note that in the ADNI and NACC study cohort,  Mini-Mental State Examination (MMSE) and Apolipoprotein E4 (APOE4) allele genetic information are unavailable for some subjects. All scans involved in this study are performed on individuals within $\pm 6$ months from the date of clinical diagnosis.

Following other AD-related studies~\cite{qiu2020development,zichao2023,Jang_2022_CVPR}, we use the ``training, internal testing, and external validation''  paradigm for experiments. In the experiment, only the MRI image from the baseline visit was selected for each subject. The internal ADNI dataset was randomly split according to the ratio of 7:1:2, where 70\% of patients are used for training, 10\%  for internal validation, and the rest for internal testing. The validation ADNI set was utilized to select hyperparameters. In addition, the external NACC dataset was used to test the generability. The MCI patients included were classified as pMCI if they progressed to AD \emph{within 36 months} and as sMCI if they did not. We highlight that during training of our \methodname, we did not use MCI subtype labels, \ie~combining sMCI and pMCI as MCI for model training.

\subsubsection{Preprocessing pipeline}
The MRI data were preprocessed through the standard preprocessing pipeline~\cite{qiu2020development,wang2024joint} and aligned to the MNI152 template space using FreeSurfer~\cite{fischl2012freesurfer} and FSL software. The preprocessing pipeline consists of the following five steps: (1) anterior commissure and posterior commissure alignment; (2) intensity correction; (3) skull stripping; (4) registration to the MNI152 template space; and (5) clipping the outliers and normalization. The first three steps were implemented via FreeSurfer. The registration was achieved via the FLIRT tool available within the FSL package. Finally, all MRI images were resized to $128\times 128\times 128$.

\subsubsection{Implementation details}
We implemented all the methods with the PyTorch library~\cite{paszke2019pytorch} and trained the networks on NVIDIA V100 GPUs. For a fair comparison, all the methods used in our experiments, including our \methodname, were built upon the RestNet18, whose learnable parameters were initialized by the Kaiming method~\cite{he2015delving}. We trained all networks 60 epochs using the Adam optimizer with $\beta_1 = 0.5$ and $\beta_2 = 0.999$. We set the initial learning rate to $2\times 10^{-4}$ and gradually reduced it using exponential decay with a decay rate $0.95$. The batch size was set to 8. 
The hyperparameter $\lamda{HybOrdinal}$ in Eq.~\eqref{equ:loss_final} was linearly increased to 1.0 as the training iteration progressed in the experiments.
For a fair and reliable performance evaluation~\cite{song2018collaborative,qian2021my}, we repeated the experiments with different random seeds \emph{five times} and reported their mean results. Five metrics were used for performance evaluation: accuracy (ACC), the area under the receiver operating characteristic curve (AUC), F1-score, precision, and recall. 

\subsubsection{Competing methods}
We conducted a comprehensive comparison of our \methodname with 18 DL-based SOTA methods. These methods can be categorized into four categories: AD-related, fine-grained, hard-soft labeled, and ordinal-based.

AD-related competing methods includes 3D-ResNet$^f_2$~\cite{korolev2017residual}, 3D-SENet~\cite{pan2020early}, I3D~\cite{carreira2017quo}, ResAttNet~\cite{zhang2021explainable}, TripleMRNet~\cite{bien2018deep}, TSM-Res18~\cite{zhang2023video4mri}, and MedicalNet~\cite{chen2019med3d}. Specifically, 3D-ResNet$^f_2$, 3D-SENet, I3D, and ResAttNet are the classical 3D classification networks for AD diagnoses very recently. 3D-SENet introduces SE~\cite{hu2018squeeze} blocks to recalibrate channel-wise features adaptively. I3D employs convolutional and pooling kernels of different sizes to acquire features across various scales. ResAttNet introduces the self-attention module to extract global features. TripleMRNet and TSM-Res18 are two recently well-performing 2D convolution-based algorithms for AD diagnosis. MedicalNet employs 3D ResNet18 as the backbone and trained it on 23 different medical datasets.

\begin{table*}[t]
	\centering
	\caption{Performance comparison  of our \methodname with other 18 SOTA competing methods. \textbf{I}, \textbf{II}, \textbf{III}, and \textbf{IV} represent AD-related, fine-grained, hard-soft labeled, and ordinal-based methods, respectively. The best results are highlighted in \textbf{bold} font.}
	\label{tab:comparison}
       \resizebox*{1\linewidth}{!}{
	\begin{tabular}{rlcccccccccccccc}
	\shline
         \multicolumn{3}{c}{\multirow{2}{*}{\textbf{Method}}}
         &\multirow{2}{*}{\tabincell{c}{\textbf{Training}\\\textbf{labels}}}
        &&\multicolumn{5}{c}{\textbf{\emph{Internal} ADNI testing set}}
         &&\multicolumn{5}{c}{\textbf{\emph{External} NACC testing set}}
          \\ \cline{6-10}\cline{12-16}
          &&&&& \textbf{ACC} &\textbf{AUC} &\textbf{F1-score}
        &\textbf{Precision} &\textbf{Recall} & &\textbf{ACC} &\textbf{AUC} &\textbf{F1-score}
        &\textbf{Precision} &\textbf{Recall}  \\
        \cline{1-4}\cline{6-10}\cline{12-16}
       1&3D-ResNet$^f_2$~\cite{korolev2017residual} &
       \multirow{7}{*}{\textbf{\large{I}}}&sMCI/pMCI&&62.4&60.9&61.6&60.3&62.4&&54.3&55.0&54.1&56.0&52.5 \\
      2&3D-SENet~\cite{pan2020early}&&sMCI/pMCI &&65.7&65.6&65.3&66.9&63.7 &&56.5&57.5&55.6&55.4&55.8 \\
      3&I3D~\cite{carreira2017quo}&&sMCI/pMCI &&66.3&66.2&66.1&67.3&65.0 &&56.8&57.8&55.5&54.1&57.0\\
      4&ResAttNet~\cite{zhang2021explainable}&&sMCI/pMCI &&66.4&66.3&65.8&66.8&64.4 &&57.2&58.4&57.0&56.6&57.2 \\
      5&TripleMRNet~\cite{bien2018deep}&&sMCI/pMCI &&65.7&65.8&65.0&65.1&64.7 &&60.8&60.3&59.0&59.7&58.3 \\
      6&TSM-Res18~\cite{zhang2023video4mri}&&sMCI/pMCI &&66.6&66.9&65.5&66.3&64.6 &&62.0&61.1&60.5&60.0&60.9 \\
      7&MedicalNet~\cite{chen2019med3d}&&sMCI/pMCI &&67.5&68.2&67.5&67.0&67.1 &&63.9&62.3&63.6&64.3&63.4 \\
            \cline{1-4}\cline{6-10}\cline{12-16}
        8&
        ProtoTree~\cite{nauta2022neural} &\multirow{4}{*}{\textbf{\large{II}}}
        &sMCI/pMCI
        &&66.4 &67.0 &66.2 &67.0 &65.4 
        &&61.8 &61.4 &61.3 &61.7 &60.9\\
        9&SAM~\cite{shu2022improving} &&sMCI/pMCI
        &&67.3&67.2&66.8
        &67.4&66.2&
        &63.0&62.6&61.8
        &62.4&61.0\\
        10&ViT-NeT~\cite{kim2022vit} &&sMCI/pMCI
        &&65.7&64.7&64.8
        &65.4&64.2
        &&56.8&57.7&56.7
        &56.8&57.2\\
        11&MPNCOV+SEB~\cite{song2022eigenvalues}&&sMCI/pMCI
        &&68.2&68.0&67.8
        &67.0&68.6
        &&63.9&62.5&62.7
        &63.4&62.2\\
         \cline{1-4}\cline{6-10}\cline{12-16}
      12&WSL~\cite{zhou2020rethinking} &\multirow{2}{*}{\textbf{\large{III}}}&NC/AD$+$sMCI/pMCI
        &&69.3&69.5&69.7
        &69.9&69.3&
        &65.8&65.1&64.5
        &64.9&64.1 \\
        13&AST~\cite{li2022asymmetric} &&NC/AD$+$sMCI/pMCI
        &&68.8&69.0&69.2
        &69.5&68.9&
        &65.2&64.4&64.2
        &64.6&63.8 \\
      \cline{1-4}\cline{6-10}\cline{12-16}
      14
      &3D-ResNet$^c_2$~\cite{huang2019diagnosis}&
      \multirow{5}{*}{\textbf{\large{IV}}}
      &NC/AD
      &&64.4&65.7&65.6
      &67.6&64.4&&57.1&58.2&56.3&58.4&54.3 \\
      15&3D-ResNet$_3$~\cite{korolev2017residual} &&NC/MCI/AD &&65.0&66.2&65.8
      &67.6&64.7&&58.1&59.3&58.0&58.5&57.4 \\
     16&OR-CNN~\cite{niu2016ordinal} &&NC/MCI/AD
      &&65.7&66.9&66.6&69.0&64.9&&61.7&61.4&60.8&60.7&60.8\\
      17&ADRank~\cite{qiao2022ranking} &&NC/MCI/AD
      &&67.0&67.9&67.4&68.4&65.7&&63.4&63.0&62.1&63.0&60.9\\
      18&CORF~\cite{zhu2021convolutional} &&NC/MCI/AD
      &&67.8&68.1&68.1&69.0&67.2&&64.1&63.5&62.8&63.2&62.1\\
        &\methodname (\textbf{ours}) &
         &NC/MCI/AD
        &&\best{72.0}&\best{71.3}&\best{71.1}&\best{71.3}&\best{71.0}
        &&\best{67.6}&\best{66.8}&\best{66.4}&\best{66.1}&\best{67.2}\\
        \shline
		\end{tabular}
 }

\end{table*}

For fine-grained methods, ProtoTree~\cite{nauta2022neural}, SAM~\cite{shu2022improving}, ViT-NeT~\cite{kim2022vit}, and MPNCOV+SEB~\cite{song2022eigenvalues} were selected as baselines. The first three fall within the  recognition of localization-classification subnetworks, whereas the last one is categorized under recognition by end-to-end feature encoding. Note that all AD-related and fine-grained competing methods are trained with sMCI and pMCI labeled data.

Furthermore, we also compared our \methodname with two recently published SOTA hard-soft labeled methods, \ie~WSL~\cite{zhou2020rethinking} and AST~\cite{li2022asymmetric}. Concretely, we first trained the 3D ResNet34 with NC and AD data as the teacher network and then used it to teach a 3D ResNet18 student network by knowledge distillation~\cite{hinton2015distilling} to differentiate sMCI and pMCI.

Moreover, we compared our \methodname with five ordinal-based methods. These include two variants of 3D ResNet trained with the CE loss on NC/AD and NC/MCI/AD labeled data, \ie~3D-ResNet$^c_2$ and 3D-ResNet$_3$. Additionally, we compared against OR-CNN~\cite{niu2016ordinal,rosati2022novel}, ADRank~\cite{qiao2022ranking}, and CORF~\cite{zhu2021convolutional}. OR-CNN transforms the ordinal classification problem into a series of binary sub-problems to make the predicted label space ordinal. ADRank adds the pairwise ranking constraint to OR-CNN through an additional FC layer. CORF integrates ordinal classification and differentiable decision trees to obtain more stable ordinal relationships. Of note, the latter three methods were trained with NC/MCI/AD labeled data and without using the MCI subtype labels, \ie~combining sMCI and pMCI as MCI for training. To ensure a fair evaluation, although many ordinal-based methods were trained with different labels, we modified the inference procedures of them as ours, \ie~predicting MCI progression by comparing the feature with the learned NC and AD global prototypes, and assessed their performance on the same internal ADNI and external NACC testing sets with sMCI/pMCI subtype labels.

\subsection{Comparison to SOTA Methods}
\label{sec:exp_sota}
\tabref{tab:comparison} presents comparative experimental results of all methods. 
Obviously, the results show that our \methodname performs better in prediction performance and has better generalizability across internal and external datasets than other methods.

For AD-related methods, by using different attention mechanisms and different kernel sizes, 3D-SENet, I3D, and ResAttNet perform better than basic 3D-ResNet$^f_2$ but are all prone to overfitting due to large parameters with 3D convolution. The performance tested in the external dataset drops quickly unless the network (\ie~MedicalNet) is pre-trained with many datasets to maintain good generalization. Moreover, 2D convolution-based methods outperform some 3D convolutional methods and have better generability. TripleMRNet and TSM-Res18 can learn more stable feature representations and present better generability than others. 

For fine-grained methods, SAM outperforms other localization-classification subnetwork methods due to its ability to regulate the network, focusing on key regions shared across different instances and classes. Moreover, MPNCOV+SEB outperforms other fine-grained methods, attributed to its emphasis on high-order statistics, enabling better modeling of subtle inter-class details.

For hard-soft labeled methods, incorporating an additional teacher network with larger parameters, along with a knowledge distillation strategy, validates beneficial to help the network to enlarge the inter-class variations, leading to improved performance. Nevertheless, these methods still face challenges in effectively learning the ordinal nature of AD.

For ordinal-based methods, the comparison between 3D-ResNet$^c_2$ and 3D-ResNet$^f_2$ shows that the 3D-ResNet$_2$ learns more helpful information from NC/AD than sMCI/pMCI, which is consistent with prior works~\cite{falahati2014multivariate,huang2019diagnosis,kwak2021subtyping,kwak2022identifying}. In addition, although 3D-ResNet$_3$ only introduces MCI subjects, it makes the inter-class variations larger. OR-CNN improves prediction performance by ordering the predicted label space. However, it cannot guarantee that the learned features are ordered. ADRank adds the additional pairwise ranking constraint in the feature space to OR-CNN, but the ranking operation is still achieved through the FC layer.
CORF leverages differentiable decision trees to divide the entire feature into several sub-features, resulting in a more stable ordinal feature space to improve performance. 

\begin{figure*}[t]
\centering
\subfigure[]{
\label{fig:gradcam_g}
\includegraphics[width=0.48\linewidth]{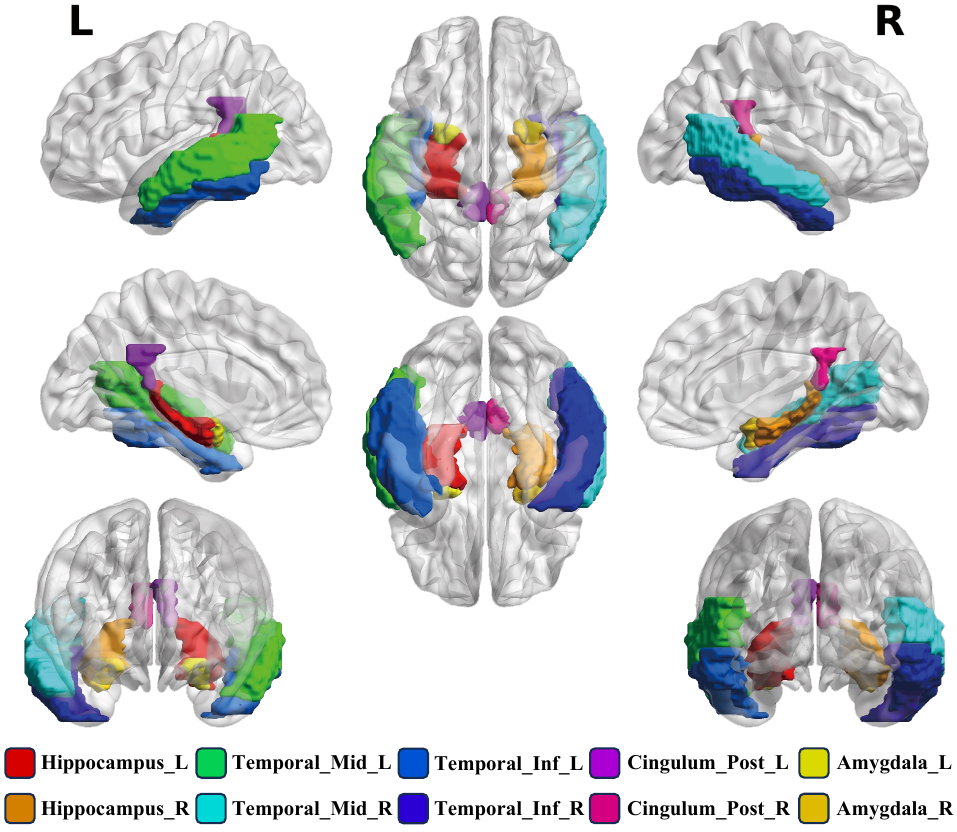}
    }
\subfigure[]{
\label{fig:gradcam_i}
\includegraphics[width=0.48\linewidth]{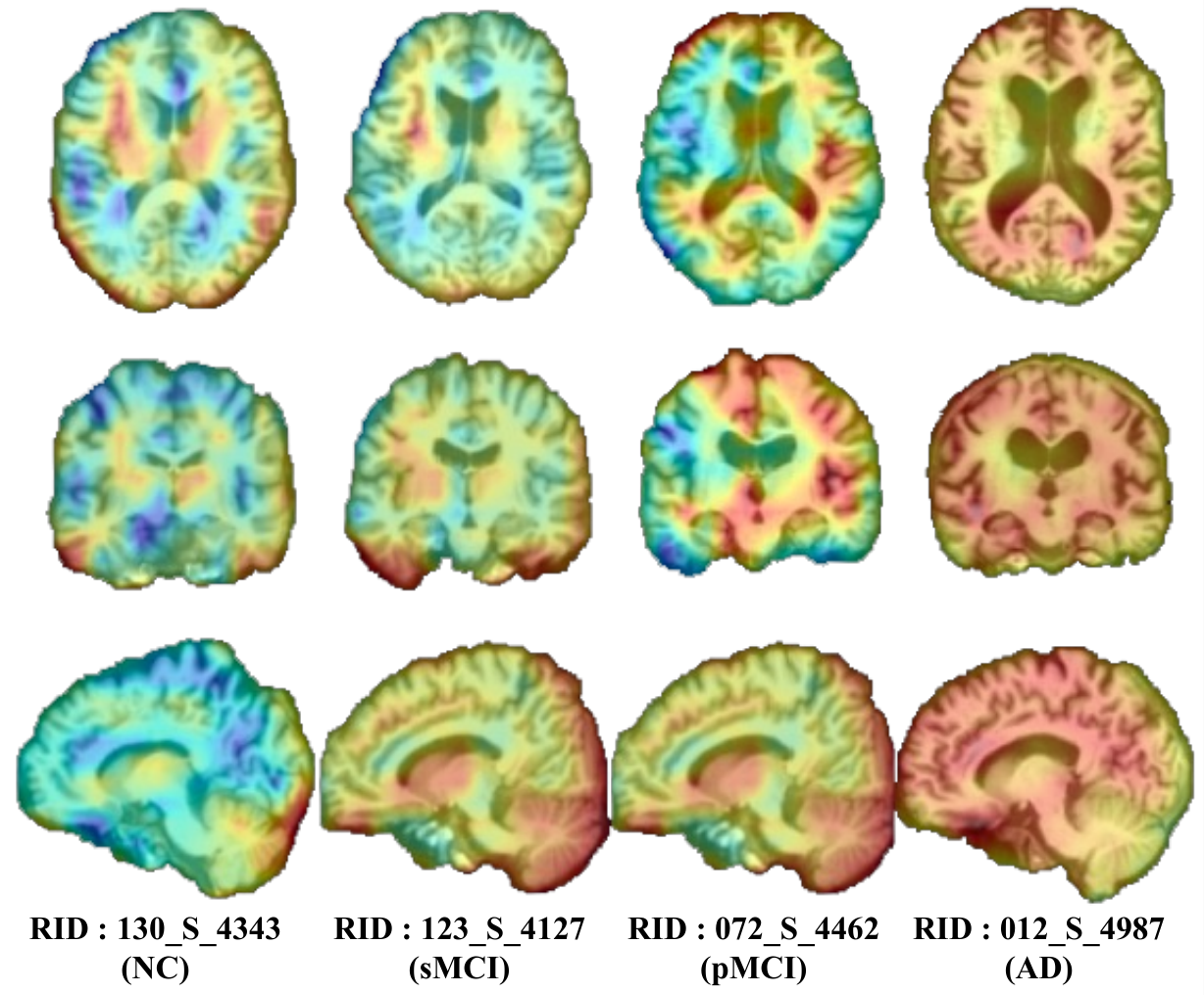}
    }
\caption{Interpretable Grad-CAM results of the proposed \methodname. (\textbf{a}) Top 10 prediction discriminative regions of interest focused by our \methodname on the internal ADNI sMCI and pMCI testing set; the capital letters L and R of a brain region label refer to the left and right cerebral hemispheres, respectively. (\textbf{b})~Grad-CAM visualization results of four randomly selected brains from all AD stages on the internal ADNI testing set.}
\label{fig:gradcam}
\end{figure*}

In summary, the results in~\tabref{tab:comparison} demonstrate that (1) leveraging coarse-grained labels can consistently improve the model performance, and (2) our \methodname consistently outperforms other methods, further confirming the effectiveness and the robustness of our approach. 
Furthermore, we highlight that our \methodname can (1) guarantee a consistent order between each low-dimensional feature space and the ordinal label space across different granularities, and (2) attain high intra-class compactness and inter-class separability for the learned features. Please refer to \ref{sec:cross} for cross-validation results. 

We also conducted a significant test between our \methodname and other competing methods through a one-sided Mann-Whitney U test~\cite{nachar2008mann} over predicted probabilities on the internal ADNI testing set, which show that all the $p$-values obtained in Table~\ref{tab:pvalue}  are below the significance level of $0.05$, suggesting that the our \methodname performs significantly better than others.

\subsection{Interpretable Visualization}
\label{sec:interpret}

Here, we employ Gradient-weighted Class Activation Mapping (Grad-CAM)~\cite{selvaraju2017grad} to present the interpretable visualization results in \figref{fig:gradcam}. First, we calculate the mean Grad-CAM value on the internal ADNI sMCI and pMCI testing set and divide the results by the brain automatic anatomical labeling atlas 3~\cite{rolls2020automated}. Fig.~\ref{fig:gradcam_g} presents the top 10 prediction regions of interest identified by our \methodname, including the hippocampus~\cite{fjell2014normal}, middle temporal gyrus (Temporal\_Mid)~\cite{convit2000atrophy}, inferior temporal gyrus (Temporal\_Inf)~\cite{chan2001patterns}, posterior cingulate cortex (Cingulum\_Post)~\cite{minoshima1997metabolic}, and amygdala, whose changes have been validated to link with the worsening of AD. In addition, the functions related to these identified brain regions are mainly involved with the behavioral domains of memory, emotion, language, perception, and activity, which are also generally consistent with the common symptoms of AD.

Moreover, we randomly select four subjects from different stages of AD on the internal ADNI testing set and present the interpretable Grad-CAM results in \figref{fig:gradcam_i}. We observed that as AD progressing, the discriminative regions of interest identified by our \methodname gradually expand in scope and intensify in significance, proving that our \methodname could help the network learn the ordinal nature of AD.

\subsection{Ablation Study}
\label{sec:ablation}
In this subsection, we first perform ablation on different loss components to demonstrate the effectiveness of the proposed all granular ordinal losses, and then study the effects of the EMA prototype update strategy.

\begin{table}[t]
\centering
\caption{Ablation results on different loss components for modernizing our \methodname on the Internal ADNI testing set. The momentum decay $\sigma$ of our EMA prototype update strategy is set to $0.9$. The best results are highlighted in \textbf{bold} font.}
\label{tab:ablation_loss}
\begin{tabular*}{1\linewidth}{@{\extracolsep{\fill}}lccccc}
\shline
\textbf{Variants} & \textbf{ACC} &\textbf{AUC} &\textbf{F1-score}&\textbf{Precision} &\textbf{Recall}
\\
\cline{1-1}\cline{2-6}
$\loss{CE}$&65.0&66.2&65.8&67.6&64.7\\
\quad$+\loss{Ins2Ins}$&68.1&68.0&67.8&68.0&67.7\\
\quad\quad$+\loss{Ins2Cls}$&70.6&69.8&70.5&70.9&70.3\\
\quad\quad\quad$+\loss{Cls2Cls}$&\best{72.0}&\best{71.3}&\best{71.1}&\best{71.3}&\best{71.0}\\
\shline
\end{tabular*}
\end{table}

\begin{figure*}[t]
\centering  
\includegraphics[width=1\textwidth]{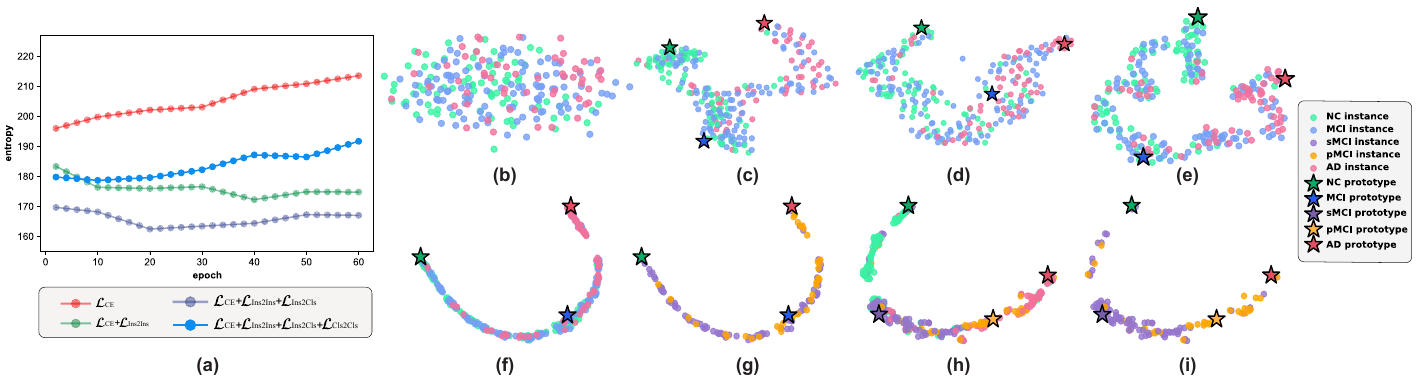}
\caption{Effectiveness of different loss components for modernizing our \methodname by entropy result presentation of the feature space and data visualization using $t$-SNE on the internal ADNI testing dataset. (\textbf{a}) Entropy results in the feature space for modernizing our \methodname by different loss components. A series of two-dimensional plots generated by $t$-SNE with the following inputs: (\textbf{b}) voxel-level MRI intensity values, (\textbf{c}) encoder outputs constrained by $\loss{CE}$ loss, (\textbf{d}) encoder outputs constrained by additional $\loss{Ins2Ins}$ loss, (\textbf{e}) encoder outputs constrained by additional $\loss{Ins2Cls}$ loss, (\textbf{f}) encoder outputs constrained by additional $\loss{Cls2Cls}$ loss with NC/MCI/AD subjects, and (\textbf{g}) encoder outputs constrained by additional $\loss{Cls2Cls}$ loss with MCI subjects marked by exact subtype labels.  (\textbf{h}) encoder outputs trained by whole four labels, \ie~NC$\rightarrow$sMCI$\rightarrow$pMCI$\rightarrow$AD. (\textbf{i}) encoder outputs trained by whole four labels with MCI subjects marked by exact subtype labels. Of note, the prototypes presented here are the global prototypes generated by our online EMA prototype update strategy.}
\label{fig:tsne}
\end{figure*}

\subsubsection{Ablation on different loss components}
Here, we present detailed quantitative ablation results with different loss components, \ie~$\loss{CE}$, $\loss{Ins2Ins}$, $\loss{Ins2Cls}$, and $\loss{Cls2Cls}$, to modernize the proposed \methodname on the internal ADNI testing set in \tabref{tab:ablation_loss}. Quantitative ablation results show that all proposed granular ordinal losses can improve prediction performance. 

Following~\cite{zhang2022improving}, we used the meanNN entropy estimator~\cite{faivishevsky2008ica} to estimate the entropy in the high-dimensional feature space trained by different loss components; the entropy of a random variable can be loosely defined as the amount of ``information'' associated with that variable, reflecting its uncertainty. Moreover, we present a series of 2-dimensional visualization plots generated by $t$-distributed stochastic neighbor embedding ($t$-SNE)~\cite{van2008visualizing} on the internal ADNI testing set to demonstrate the effectiveness of different loss components in \figref{fig:tsne}.

\figref{fig:tsne}(b) visualizes the volumetric 3D T1 MRI scans using the intensity values as inputs, which reveals no clear differentiation in NC, MCI, and AD. \figref{fig:tsne}(c) visualizes the outputs of the encoder $f_{\vct{\theta}}(\cdot)$ trained with only $\loss{CE}$, which can generate a high entropy feature space, allowing for clustering the data into three different stages roughly, but lag in the ability to capture the inherent ordinality of AD. Then, we showed the outputs of the encoder trained with the additional instance-to-instance loss $\loss{Ins2Ins}$ in \figref{fig:tsne}(d). We observed that the encoder trained with $\loss{Ins2Ins}$ can learn the ordinal development of AD, but the decision boundary among different classes becomes ambiguous. In contrast, the introduction of the additional instance-to-class loss $\loss{Ins2Cls}$ can make the learned feature high intra-class compactness, and the $t$-SNE result is shown in \figref{fig:tsne}(e). However, although $\loss{Ins2Ins}$ and $\loss{Ins2Cls}$ can help the encoder learn the ordinal nature of AD to aid in MCI progression prediction,  we found that they substantially decreased the entropy of the feature space, as shown in \figref{fig:tsne}(a). Furthermore, our HOPE leveraged the proposed class-to-class loss $\loss{Cls2Cls}$ to increase the inter-class separability of the learned features and reinforce the ordinal nature based on the progressive development of AD in \figref{fig:tsne}(f-g). We observed that the encoder trained with additional $\loss{Cls2Cls}$ can increase the entropy of the feature space by enlarging the distances among feature representations while weighting the distances to preserve the ordinal relationship. When we marked finer-grained MCI labels in \figref{fig:tsne}(g), we found that the sMCI and pMCI features are closer to NC and AD prototypes, respectively.

Regarding the phenomenon that the same class does not get together, we identify three key factors. 
\textbf{(i)} \emph{Ordinal feature learning in our \methodname}. The proposed \emph{hybrid-granularity ordinal loss} can help our \methodname learn an ordinal metric space for predicting MCI progression, rather than just learning a discriminative feature space like conventional classification problems~\cite{li2021learning,zhu2020ordinal,yu2023retinal}. 
\textbf{(ii)} \emph{Ambiguous decision boundary among AD subtypes}. The decision boundary among different AD stages' subjects is more ambiguous than natural image-based classification problems. For example, according to the ADNI differentiate criterion~\cite{aisen2010clinical}, the Mini-Mental State Examination (MMSE) score ranges of NC, MCI, and AD are [24-30], [24-30], and [20-26], respectively, and there obviously exists overlaps among different stages. 
\textbf{(iii)} \emph{Heterogeneity of MRI scans}. The ADNI structural MRI data are acquired from 59 different sites using various scanning vendors, such as GE, Philips, and Siemens, and different magnetic strengths, such as 1.5 / 3 Tesla, and even from different countries~\cite{aisen2010clinical, qiu2020development}. Please refer to \ref{sec:more_result} for 
more explanations about our $t$-SNE results. 

\begin{table}[t]
\centering
\caption{Ablation results of our online EMA prototype update strategy on the internal ADNI validation set. The best results are highlighted in \textbf{bold} font.}
\label{tab:ablation_mome_decay}
\begin{tabular*}{1\linewidth}{@{\extracolsep{\fill}}clccccc}
\shline \textbf{EMA}
&
$\sigma$ & \textbf{ACC} &\textbf{AUC} &\textbf{F1-score}&\textbf{Precision} &\textbf{Recall}
\\
\cline{1-2}\cline{3-7}
\XSolidBrush&-&70.0&69.4&69.1&68.8&69.5\\
\CheckmarkBold&0.5&70.4&69.6&69.8&69.0&70.6\\
\CheckmarkBold&0.8&71.3&70.4&70.5&69.9&70.9\\
\CheckmarkBold&0.9&\best{72.3}&\best{71.9}&\best{71.7}&\best{71.2}&\best{72.0}\\
\CheckmarkBold&0.99&71.9&71.1&70.9&70.6&71.3\\
\CheckmarkBold&0.999&70.9&70.3&70.0&69.6&70.6\\
\shline
\end{tabular*}
\end{table}

\subsubsection{Ablation on the EMA prototype update strategy} 
We performed an ablation study of the EMA prototype update strategy on the internal ADNI validation sMCI and pMCI set, and the results are shown in \tabref{tab:ablation_mome_decay}.
All experiments used $\loss{Total}$ in Eq.~\eqref{equ:loss_final} to optimize the network. We would like to note that when the EMA update strategy was not used, the global prototypes were calculated similarly to \equref{equ:mean value1} by feeding the corresponding stages' data from the training set into the network trained in the last iteration.

Concretely, we have the following observations from \tabref{tab:ablation_mome_decay}. \textbf{(i)} The use of the EMA update strategy makes the global prototype learning more stable, resulting in improved performance compared to not employing this strategy. In addition, once the training of the proposed \methodname is completed, the inference speed of the proposed \methodname is as fast as that of the conventional network based on the fully-connected layer without extra prototype calculating time. 
\textbf{(ii)} Setting the momentum decay $\sigma$ too low or too high adversely affected the prediction performance. Finally, our \methodname adopts a value of 0.9 as the momentum decay parameter in our online EMA prototype update strategy.

\subsection{Extension}
\label{sec:fine-tune}

For the case where substantial subtype-labeled data (sMCI and pMCI) are available, we further modified our \emph{hybrid-granularity ordinal loss} to facilitate learning the ordinal nature of all four labels (NC$\rightarrow$sMCI$\rightarrow$pMCI$\rightarrow$AD).  Specifically, we have two variants of our \methodname: (1) V1 is the original \methodname trained with three coarse-labeled data, \ie~NC$\rightarrow$MCI$\rightarrow$AD,
(2) V2 is the modified \methodname trained with four labeled data, \ie~NC$\rightarrow$sMCI$\rightarrow$pMCI$\rightarrow$AD. Note that V2 conducts inference by comparing the to-be-predicted MCI feature with the learned sMCI and pMCI global prototypes.

\begin{table}[!t]
\centering
\caption{Extension results of our \methodname trained with three (NC, MCI, and AD) and four (NC, sMCI, pMCI, and AD) labeled data.
}
\label{tab:modified_4}
\resizebox*{1\linewidth}{!}{
\begin{tabular}{lcccccc}
\shline
\textbf{Variants} &\tabincell{c}{\textbf{Training}\\\textbf{labels}}&\textbf{ACC}&\textbf{AUC} &\textbf{F1-score}&\textbf{Precision} &\textbf{Recall}
\\
\cline{1-2}\cline{3-7}
\textbf{V1}&NC/MCI/AD&72.0&71.3&71.1&71.3&71.0\\
\textbf{V2}&NC/sMCI/pMCI/AD&\textbf{73.0}&\textbf{73.1}&\textbf{73.0}&\textbf{73.1}&\textbf{72.8}\\
\shline
\end{tabular}
}
\end{table}

\tabref{tab:modified_4} presents the  quantitative results and \figref{fig:tsne}~(h-i) shows the $t$-SNE results.
 We can see that V2 can achieve better performance and generate a more ordinal and differentiated feature space. Looking from an alternative standpoint, our original \methodname, trained with coarse-grained labels (\ie~V1), still demonstrates satisfactory performance, affirming its applicability to real-world scenarios.

\section{Discussion}
\label{sec:dicussion}
This section delves into the discussion of related works, highlighting the advantages of our \methodname, its applicability to other domains, and acknowledging its limitations.

\noindent\textbf{Discussion on the related work}\quad Here, we discuss the relationship of our \methodname and other fine-grained, hard-soft labeled, and ordinal-based studies. 

\textbf{(i)} Compared to fine-grained studies, distinguishing pMCI from sMCI also has a similar problem with other fine-grained classification tasks, \ie~\emph{small inter-class variations} caused by highly similar subcategories, and the \emph{large intra-class variations} in individual differences~\cite{yang2018learning,wei2021fine}. Unlike other fine-grained classification methods that mine the fine-grained discriminative information among subcategories (sMCI and pMCI), our \methodname takes the ordinal nature of AD into account and uses larger inter-class variations among coarse-grained labels (NC, MCI, and AD) to distinguish them.

\textbf{(ii)} Compared to hard-soft labeled methods, conventional methods improve performance by enlarging inter-class variations, which, unfortunately, remain dependent on subtype-labeled data and
are unable to capture the ordinal nature of AD. The comparative results in~\tabref{tab:comparison} also suggest that although aided by the higher parameters teacher network, hard-soft labeled methods still perform worse than ours. 

\textbf{(iii)} Compared to ordinal-based methods, the biggest weakness of existing ordinal learning methods is that most of them focus on either (1)  making the predicted label space ordinal~\cite{niu2016ordinal,zhu2021convolutional,lei2022meta} or (2)  constraining the feature space to satisfy particular distributions, \eg~Poisson or Bernoulli distribution~\cite{wang2023controlling}.
However, these methods pose challenges to handling disease-related problems. For example, it is uncertain if the concrete transition period between discrete disease stages NC$\rightarrow$MCI equals MCI$\rightarrow$AD, or if the feature space adheres to a particular assumed distribution. The only thing we know about disease-related problems is the order of the disease progression. To address these weaknesses, we design a novel hybrid-granularity ordinal loss to (1) directly rank different subjects to be ordinal in the feature space, and (2) avoid predefined distribution assumption and only rely on the prior knowledge of the disease progression order to rank different subjects, which are more faithful to the AD clinical knowledge and can learn the ordinal nature of AD better.

\noindent\textbf{Discussion on the advantages}\quad This paper proposed a novel hybrid-granularity ordinal prototype learning method for the progression prediction of MCI, with the advantages of our \methodname being highlighted as follows. \textbf{(i)} \emph{Closely integrated with clinical knowledge and interpretability.} Existing MCI progression prediction works typically require finer-grained MCI subtype labels (\ie~sMCI and pMCI). However, we regard the extensive cross-sectional data collected from subjects with different coarse-grained diagnosis labels, \ie~NC$\rightarrow$MCI$\rightarrow$AD, as the ``latent'' longitudinal data specific to the entire disease duration of the AD cohort. Towards that end, we introduce the prior knowledge of disease progression into our \methodname to learn the ordinal nature of AD.  Moreover, the visualization results using $t$-SNE and Grad-CAM also show that our \methodname has made the learned feature ordinal across hybrid granularities, high intra-class compactness, and inter-class separation. \textbf{(ii)} \emph{Better prediction performance and generability.} Extensive experimental results on the internal ADNI and external NACC datasets show that our \methodname outperforms recently published SOTA DL-based methods and has better generalizability. They also show that predicting MCI progression using extensive cross-sectional data from different coarse-grained diagnosis labels, \ie~NC, MCI, and AD, is flexible and performs better than using MCI subtype labels, \ie~sMCI and pMCI.
\textbf{(iii)} \emph{More applicable for real scenarios.} Although most ADNI structural MRI data have sub-labels of sMCI or pMCI, the heterogeneity of ADNI MRI scans sets it apart from MRI data collected at a specific real-world site. This heterogeneity adversely affects the generalization performance on external sites, particularly for methods that solely rely on sub-labels. For the data collected from one specific site, we highlight that the coarse-grained data is \emph{more accessible} and \emph{sufficient}, affirming our \methodname is  well-suited for practical applications. Moreover, when substantial subtype-labeled data (sMCI and pMCI) are available, our \methodname can be further extended for improved performance.

\noindent\textbf{Discussion on the applicability to other domains}\quad We highlight that our \methodname can be applied effectively to other domains, such as progressive disease classification and ordinal-based uncertainty problems. \textbf{(i)} Regarding progressive diseases like Parkinson's disease, Huntington's disease, multiple sclerosis, and various types of cancer, predicting the deterioration of patients in the transitional phase---the phase between normal and severe stages---holds significant relevance for subsequent interventions. It is worth noting that our \methodname is also applicable in predicting the progression of these diseases due to its ordinal-aware design. \textbf{(ii)} Regarding ordinal-based uncertainty problems, \eg~the tumor segmentation task that involves annotations from multiple radiologists~\cite{chen2023berdiff, schmidt2023probabilistic}, it is common to observe consistent annotations for extremely normal and abnormal pixels. However, the real challenge lies in identifying the uncertain pixels, where inconsistent annotations may arise among radiologists. Our \methodname also offers valuable insight into tackling the segmentation of uncertain pixels by learning the ordinal annotation standard, \ie~consistently labeled normal pixels $\rightarrow$ inconsistently labeled / uncertain pixels $\rightarrow$ consistently labeled abnormal pixels, for different radiologists.

\noindent\textbf{Discussion on the limitations}\quad Here, we acknowledge some limitations in this work. \textbf{(i)} Considering the high cost associated with prospectively acquiring longitudinal data, we leverage the extensive cross-sectional data for MCI progression prediction. However, how to effectively combine longitudinal data with the ordinal nature of AD is a valuable research direction. \textbf{(ii)} Although the proposed \methodname performs well and has better generalizability than compared methods, it cannot predict the exact time to progress to AD for pMCI; we will focus on this problem and integrate exact time into the model to improve prediction performance in the future.
\section{Conclusion}
\label{sec:conclusion}
In this paper,  we proposed a novel hybrid-granularity ordinal prototype learning method to distinguish pMCI from sMCI by learning the ordinal nature of AD, yielding a more applicable, generalizable algorithm towards real-world scenarios. The proposed hybrid-granularity ordinal loss effectively integrates instance-to-instance ordinality, instance-to-class compactness, and class-to-class separation. The introduced EMA update strategy can make the prototype learning more stable. Extensive experimental results on the internal ADNI and external NACC datasets validated the effectiveness of the proposed \methodname,  showcasing superior generalizability and interpretability compared to existing SOTA methods.

In the future, we will focus on leveraging the longitudinal data from each subject and predicting the exact progression time for pMCI.


\clearpage 
\setcounter{figure}{0}
\setcounter{section}{0}
\setcounter{table}{0}
\renewcommand{\thefigure}{A\arabic{figure}}
\renewcommand{\thesubfigure}{(\alph{subfigure})}
\renewcommand{\theequation}{A\arabic{equation}}
\renewcommand{\thesection}{Appendix \Roman{section}}
\renewcommand{\thetable}{A-\Roman{table}}


\section{Cross-validation Results}
\label{sec:cross}

We would like to note that we followed other AD-related studies~\cite{qiu2020development,zichao2023,Jang_2022_CVPR} to use the ``training, internal testing, and external validation''  paradigm for experiments.

\subsection{Cross-validation setting}
Here, we conducted 5-fold cross-validation for our \methodname on the Internal ADNI dataset as shown in \tabref{tab:cross_valid}. 
Specifically, we further split our training + validation data (80\%) into four folds, with our testing set (20\%) as the fifth fold.

\subsection{Competing cross-validation results}
For a better comparison, we also re-run \textbf{Top-5} competing methods for cross-validation, to further validate the effectiveness and the robustness of our \methodname as shown in \tabref{tab:cross_compt}. 

Additionally, we would like to clarify that 
CORF~\cite{zhu2021convolutional}, WSL~\cite{zhou2020rethinking}, AST~\cite{li2022asymmetric}, and our \methodname utilized the same NC and AD cohorts for different fold experiments. However, there is no overlap of MCI cohorts within each fold. The results in Tables~\ref{tab:cross_valid} and~\ref{tab:cross_compt} demonstrate that (1) leveraging coarse-grained labels can consistently improve the model performance, and (2) our \methodname consistently outperforms other methods, further confirming the effectiveness and the robustness of our approach.

\begin{table}[!h]
\centering
\caption{Cross-validation results of our \methodname on the Internal ADNI dataset.}
\label{tab:cross_valid}
\begin{tabular*}{1\linewidth}{@{\extracolsep{\fill}}cccccc}
\shline
\textbf{Fold} & \textbf{ACC} &\textbf{AUC} &\textbf{F1-score}&\textbf{Precision} &\textbf{Recall}
\\
\cline{1-1}\cline{2-6}
\textbf{1}&71.7&71.2&71.3&71.3&71.4\\
\textbf{2}&71.0&70.8&71.1&71.2&71.0\\
\textbf{3}&72.6&71.8&72.4&72.6&72.1\\
\textbf{4}&72.8&72.7&72.9&73.0&72.8\\
\textbf{5}&72.0&71.3&71.1&71.3&71.0\\
\textbf{Overall}&72.0$\pm$0.6&71.6$\pm$0.7&71.7$\pm$0.7&71.9$\pm$0.8&71.7$\pm$0.7\\
\shline
\end{tabular*}
\end{table}

\begin{table}[!h]
\centering
\caption{Comparative
cross-validation results of Top-5 competing methods and ours on the internal ADNI testing set. The best and second best results are highlighted in \textbf{bold} and \underline{underline} fonts, respectively. 
}
\label{tab:cross_compt}
\resizebox*{1\linewidth}{!}{
\begin{tabular}{lccccc}
\shline
\textbf{Methods} & \textbf{ACC} &\textbf{AUC} &\textbf{F1-score}&\textbf{Precision} &\textbf{Recall}
\\
\cline{1-1}\cline{2-6}
MedicalNet~\cite{chen2019med3d}&67.9$\pm$1.2&68.0$\pm$1.0&67.3$\pm$1.4&67.4$\pm$1.1&67.0$\pm$0.8 \\
CORF~\cite{zhu2021convolutional}&68.5$\pm$0.7&68.4$\pm$0.9&68.6$\pm$0.8&68.6$\pm$0.8&68.4$\pm$0.7\\
MPNCOV+SEB~\cite{song2022eigenvalues}&68.1$\pm$0.9&68.1$\pm$1.0&68.2$\pm$1.0&68.5$\pm$0.9&68.0$\pm$1.1\\
WSL~\cite{zhou2020rethinking}&\underline{70.2$\pm$0.7}&\underline{70.0$\pm$0.9}&\underline{71.2$\pm$0.9}&\textbf{72.0$\pm$1.0}&\underline{70.3$\pm$0.8}\\
AST~\cite{li2022asymmetric}&69.2$\pm$0.8&69.5$\pm$0.8&69.5$\pm$0.8&70.0$\pm$0.8&69.1$\pm$0.8\\
\methodname (\textbf{ours})&\textbf{72.0$\pm$0.6}&7\textbf{1.6$\pm$0.7}&\textbf{71.7$\pm$0.7}&\underline{71.9$\pm$0.8}&\textbf{71.7$\pm$0.7}\\
\shline
\end{tabular}
}
\end{table}

\section{More Details about Significant Evaluation}
\label{sec:significant}

We conducted a significant evaluation between our \methodname and other competing methods through a one-sided Mann-Whitney U test~\cite{nachar2008mann} with respect to predicted probabilities on the internal ADNI testing set, and the $p$-values are shown in \tabref{tab:pvalue}. The results show that all the $p$-values obtained are below the significance level of $0.05$, suggesting that the our \methodname  perform significantly better than the other 18 methods.

\begin{table}[!h]
\centering
\caption{Detailed obtained p-value for all competing methods.}
\label{tab:pvalue}
\resizebox*{1\linewidth}{!}{
\begin{tabular}{llccllccllc}
\shline
\multicolumn{2}{l}{\textbf{Methods}}&\textbf{p-value}&
&\multicolumn{2}{l}{\textbf{Methods}}&\textbf{p-value}&
&\multicolumn{2}{l}{\textbf{Methods}}&\textbf{p-value}
\\
\cline{1-3}\cline{5-7}\cline{9-11}
1&\textbf{3D-ResNet$_{2}^f$}&$1.2E-5$&
&7&\textbf{MedicalNet}&$9.4E-3$&
&13&\textbf{ProtoTree}&$2.3E-3$\\
2&\textbf{3D-SENet}&$1.9E-4$&
&8&\textbf{3D-ResNet$_2^c$}&$2.5E-4$&
&14&\textbf{SAM}&$3.1E-3$\\
3&\textbf{I3D}&$5.2E-4$&
&9&\textbf{3D-ResNet$_3$}&$5.2E-4$&
&15&\textbf{ViT-NeT}&$3.6E-4$\\
4&\textbf{ResAttNet}&$5.3E-4$&
&10&\textbf{OR-CNN}&$1.8E-3$&
&16&\textbf{MPNCOV+SEB}&$9.0E-3$\\
5&\textbf{TripleMRNet}&$3.1E-4$&
&11&\textbf{ADRank}&$8.2E-3$&
&17&\textbf{WSL}&$9.9E-3$\\
6&\textbf{TSM-Res18}&$1.8E-3$&
&12&\textbf{CORF}&$9.4E-3$&
&18&\textbf{AST}&$9.6E-3$\\
\shline
\end{tabular}
}
\end{table}

\section{More Explanation about $t$-SNE Result}
\label{sec:more_result}

Here, we conducted more explanation about our $t$-SNE Results for two phenomenons: \textbf{(i)} the same class does not get together, and \textbf{(ii)} the prototype location is not in the center.

\begin{figure}[!h]
    \centering
    \includegraphics[width=1\linewidth]{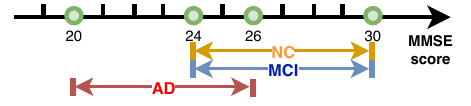}
    \caption{The Mini-Mental State Examination score ranges of different AD stages.}
    \label{fig:MMSE_range}
\end{figure}

Regarding the phenomenon that the same class does not get together, we identify three key factors:
\begin{enumerate}
    \item \emph{Ordinal feature learning}: We proposed an \emph{hybrid-granularity ordinal loss} to help our \methodname learn an ordinal metric space for predicting MCI progression, rather than just learn a discriminative feature space like conventional classification problem. This situation also occurs in related ordinal-based works~\cite{li2021learning,zhu2020ordinal,yu2023retinal}.

    \item \emph{Ambiguous decision boundary among AD subtypes}: The decision boundary between different AD stages' subjects is more ambiguous than natural image-based classification problems. For example, according the ADNI differentiate criterion~\cite{aisen2010clinical}, the Mini-Mental State Examination (MMSE) score ranges of NC, MCI, and AD are [24-30], [24-30], and [20-26], respectively, and there obviously exists overlaps between different stages, as shown in~\figref{fig:MMSE_range}.

    \item \emph{Heterogeneity of MRI scans}: The ADNI structural MRI data are acquired from 59 different sites using various scanning vendors, such as GE, Philips, and Siemens, and different magnetic strengths, such as 1.5 Tesla and 3 Tesla, and even from different countries~\cite{aisen2010clinical, qiu2020development}. 
\end{enumerate}

Regarding the phenomenon that the prototype location is deviated from the center. We would like to note that this is caused by the proposed online learning strategy. We have two ways to produce prototypes for final comparison: \textbf{(i)} \emph{Offline generation:} After the model training is done, one can calculate prototypes by averaging all the features of  samples belonging to one class in the training set. \textbf{(ii)} \emph{Online generation:} One can also produce prototypes using the proposed online EMA prototype strategy during training.

\begin{figure}[!ht]
    \centering
    \includegraphics[width=1\linewidth]{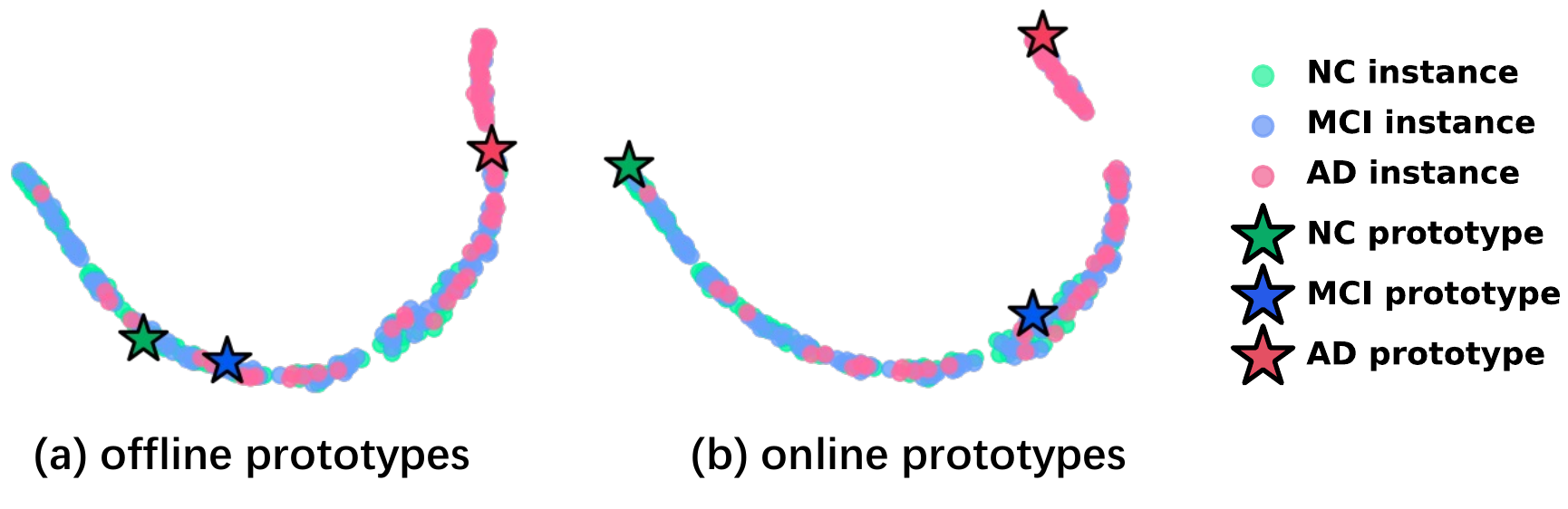}
    \caption{Data visualization using $t$-SNE on the internal ADNI testing set for our \methodname, with the global prototypes generated by offline vs. our online EMA prototype strategy.}
    \label{fig:online_revision}
\end{figure}

\begin{figure}[!ht]
    \centering
    \includegraphics[width=1\linewidth]{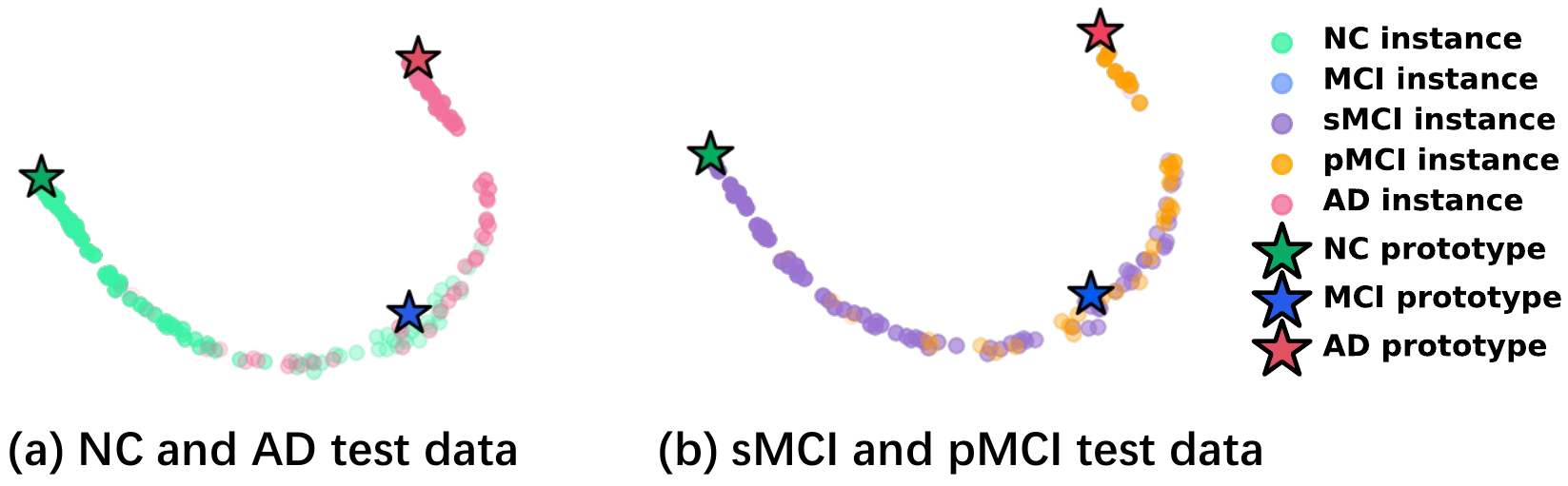}
    \caption{Data visualization using $t$-SNE on the internal ADNI testing set for our \methodname, marked by NC/AD and sMCI/pMCI, with data point transparency indicating corresponding probability.}
    \label{fig:prob_revision}
\end{figure}

\figref{fig:online_revision} presents $t$-SNE results using offline and online prototype generations for our \methodname. Note that the same $t$-SNE configuration is used. 
We have the following observations.
\begin{enumerate}
    \item \emph{One-dimensional ordinal relationship.} Our hybrid-granularity ordinal loss encourages the low-dimensional feature space to learn a one-dimensional ordinal relationship, rather than just learn a discriminative feature space like conventional classification problem. The use of $\cos$ distance in comparing prototypes and instance features results in a semicircle-like visualization, ideally extending from the value $-1$ to $1$, corresponding to the disease stages from NC to AD. Experimentally, $\cos(\boldsymbol{p}_{NC},\boldsymbol{p}_{AD}\!)\!\!=\!-0.845\!\approx\!-1$ for our online prototypes.

    \item \emph{Starting and ending point.} Owing to the continuous updates during the training process, our online EMA prototype strategy ensures that the learned NC and AD prototypes act as the starting and ending points, enhancing the quality of comparisons and leading to improved performance. Here, we also present the $t$-SNE results from \methodname, marked by NC/AD and sMCI/pMCI, with data point transparency indicating corresponding probability in \figref{fig:prob_revision}. It is observed that data points nearer to the NC and AD prototypes exhibit deeper colors, corroborating the efficacy of the prototype learning.
    
    The quantitative ablation results presented in the manuscript further demonstrate its effectiveness: this strategy achieves higher performance across all metrics compared to offline prototype generation.
\end{enumerate}

\begin{thebibliography}{10}
\providecommand{\url}[1]{#1}
\csname url@samestyle\endcsname
\providecommand{\newblock}{\relax}
\providecommand{\bibinfo}[2]{#2}
\providecommand{\BIBentrySTDinterwordspacing}{\spaceskip=0pt\relax}
\providecommand{\BIBentryALTinterwordstretchfactor}{4}
\providecommand{\BIBentryALTinterwordspacing}{\spaceskip=\fontdimen2\font plus
\BIBentryALTinterwordstretchfactor\fontdimen3\font minus \fontdimen4\font\relax}
\providecommand{\BIBforeignlanguage}[2]{{%
\expandafter\ifx\csname l@#1\endcsname\relax
\typeout{** WARNING: IEEEtran.bst: No hyphenation pattern has been}%
\typeout{** loaded for the language `#1'. Using the pattern for}%
\typeout{** the default language instead.}%
\else
\language=\csname l@#1\endcsname
\fi
#2}}
\providecommand{\BIBdecl}{\relax}
\BIBdecl

\bibitem{winblad2016defeating}
B.~Winblad \emph{et~al.}, ``Defeating {Alzheimer's} disease and other dementias: a priority for european science and society,'' \emph{Lancet Neurol.}, vol.~15, no.~5, pp. 455--532, 2016.

\bibitem{american2013diagnostic}
A.~P. Association, \emph{Diagnostic and statistical manual of mental disorders: {DSM-5}}.\hskip 1em plus 0.5em minus 0.4em\relax Am. Psychiatr. Assoc., 2013, vol.~5.

\bibitem{qiu2020development}
S.~Qiu \emph{et~al.}, ``Development and validation of an interpretable deep learning framework for {Alzheimer’s} disease classification,'' \emph{Brain}, vol. 143, no.~6, pp. 1920--1933, 2020.

\bibitem{huang2019diagnosis}
Y.~Huang \emph{et~al.}, ``Diagnosis of {Alzheimer’s} disease via multi-modality {3D} convolutional neural network,'' \emph{Front. Neurosci.}, vol.~13, p. 509, 2019.

\bibitem{pan2020early}
D.~Pan \emph{et~al.}, ``Early detection of {Alzheimer’s} disease using magnetic resonance imaging: a novel approach combining convolutional neural networks and ensemble learning,'' \emph{Front. Neurosci.}, vol.~14, p. 259, 2020.

\bibitem{beheshti2017classification}
I.~Beheshti \emph{et~al.}, ``Classification of {Alzheimer's} disease and prediction of mild cognitive {impairment-to-Alzheimer's} conversion from structural magnetic resource imaging using feature ranking and a genetic algorithm,'' \emph{Comput. Biol. Med.}, vol.~83, pp. 109--119, 2017.

\bibitem{ashtari2022multi}
M.~Ashtari-Majlan, A.~Seifi, and M.~M. Dehshibi, ``A multi-stream convolutional neural network for classification of progressive {MCI} in {Alzheimer’s} disease using structural {MRI} images,'' \emph{IEEE J. Biomed. Health Inform.}, vol.~26, no.~8, pp. 3918--3926, 2022.

\bibitem{falahati2014multivariate}
F.~Falahati \emph{et~al.}, ``Multivariate data analysis and machine learning in {Alzheimer's} disease with a focus on structural magnetic resonance imaging,'' \emph{J. Alzheimers Dis.}, vol.~41, no.~3, pp. 685--708, 2014.

\bibitem{kwak2021subtyping}
K.~Kwak, K.~S. Giovanello, A.~Bozoki, M.~Styner, and E.~Dayan, ``Subtyping of mild cognitive impairment using a deep learning model based on brain atrophy patterns,'' \emph{Cell Rep. Med.}, vol.~2, no.~12, 2021.

\bibitem{kwak2022identifying}
K.~Kwak \emph{et~al.}, ``Identifying the regional substrates predictive of {Alzheimer's} disease progression through a convolutional neural network model and occlusion,'' \emph{Hum. Brain Mapp.}, vol.~43, no.~18, pp. 5509--5519, 2022.

\bibitem{guo2021longitudinal}
T.~Guo \emph{et~al.}, ``Longitudinal cognitive and biomarker measurements support a unidirectional pathway in {Alzheimer’s} disease pathophysiology,'' \emph{Biol. Psychiatry}, vol.~89, no.~8, pp. 786--794, 2021.

\bibitem{mill2022recent}
J.~Mill and L.~Li, ``Recent advances in understanding of {Alzheimer’s} disease progression through mass spectrometry-based metabolomics,'' \emph{Phenomics}, vol.~2, no.~1, pp. 1--17, 2022.

\bibitem{jack2018nia}
C.~R. Jack~Jr \emph{et~al.}, ``{NIA-AA} research framework: toward a biological definition of {Alzheimer's} disease,'' \emph{Alzheimers. Dement.}, vol.~14, no.~4, pp. 535--562, 2018.

\bibitem{zhang2022influence}
Y.~Zhang, J.~Lu, M.~Wang, C.~Zuo, and J.~Jiang, ``Influence of gender on tau precipitation in {Alzheimer’s} disease according to {ATN} research framework,'' \emph{Phenomics}, pp. 1--11, 2022.

\bibitem{tian2023international}
M.~Tian \emph{et~al.}, ``International nuclear medicine consensus on the clinical use of amyloid positron emission tomography in {Alzheimer’s} disease,'' \emph{Phenomics}, vol.~3, no.~4, pp. 375--389, 2023.

\bibitem{vos2015prevalence}
S.~J. Vos \emph{et~al.}, ``Prevalence and prognosis of {Alzheimer’s} disease at the mild cognitive impairment stage,'' \emph{Brain}, vol. 138, no.~5, pp. 1327--1338, 2015.

\bibitem{parnetti2019prevalence}
L.~Parnetti \emph{et~al.}, ``Prevalence and risk of progression of preclinical {Alzheimer’s} disease stages: a systematic review and meta-analysis,'' \emph{Alzheimers Res. Ther.}, vol.~11, pp. 1--13, 2019.

\bibitem{chen2019med3d}
S.~Chen, K.~Ma, and Y.~Zheng, ``{Med3D}: Transfer learning for {3D} medical image analysis.'' \emph{arXiv: 1904.00625}, 2019.

\bibitem{zhang2021explainable}
X.~Zhang, L.~Han, W.~Zhu, L.~Sun, and D.~Zhang, ``An explainable {3D} residual self-attention deep neural network for joint atrophy localization and {Alzheimer’s} disease diagnosis using structural {MRI},'' \emph{IEEE J. Biomed. Health Inform.}, vol.~26, no.~11, pp. 5289--5297, 2021.

\bibitem{bien2018deep}
N.~Bien \emph{et~al.}, ``Deep-learning-assisted diagnosis for knee magnetic resonance imaging: development and retrospective validation of {MRNet},'' \emph{{PLos Med.}}, vol.~15, no.~11, p. e1002699, 2018.

\bibitem{wang2024joint}
C.~Wang \emph{et~al.}, ``Joint learning framework of cross-modal synthesis and diagnosis for alzheimer’s disease by mining underlying shared modality information,'' \emph{Med. Image Anal.}, vol.~91, p. 103032, 2024.

\bibitem{zichao2023}
Z.~Zhang, X.~Zhao, G.~Dong, and X.-M. Zhao, ``Improving {Alzheimer's} disease diagnosis with multi-modal {PET} embedding features by a 3d multi-task {MLP-Mixer} neural network,'' \emph{IEEE J. Biomed. Health Inform.}, vol.~27, no.~8, pp. 4040--4051, 2023.

\bibitem{gao2023hybrid}
X.~Gao, H.~Cai, and M.~Liu, ``A hybrid multi-scale attention convolution and aging transformer network for {Alzheimer's} disease diagnosis,'' \emph{IEEE J. Biomed. Health Inform.}, 2023.

\bibitem{wei2021fine}
X.-S. Wei \emph{et~al.}, ``Fine-grained image analysis with deep learning: A survey,'' \emph{IEEE Trans. Pattern Anal. Mach. Intell.}, vol.~44, no.~12, pp. 8927--8948, 2021.

\bibitem{yang2018learning}
Z.~Yang, T.~Luo, D.~Wang, Z.~Hu, J.~Gao, and L.~Wang, ``Learning to navigate for fine-grained classification,'' in \emph{ECCV}, 2018, pp. 420--435.

\bibitem{nauta2022neural}
M.~Nauta \emph{et~al.}, ``Neural prototype trees for interpretable fine-grained image recognition,'' in \emph{CVPR}, 2021, pp. 14\,933--14\,943.

\bibitem{shu2022improving}
Y.~Shu \emph{et~al.}, ``Improving fine-grained visual recognition in low data regimes via self-boosting attention mechanism,'' in \emph{ECCV}, 2022.

\bibitem{kim2022vit}
S.~Kim \emph{et~al.}, ``{ViT-NeT}: Interpretable vision t ransformers with neural tree decoder,'' in \emph{ICML}, 2022, pp. 11\,162--11\,172.

\bibitem{song2022eigenvalues}
Y.~Song, N.~Sebe, and W.~Wang, ``On the eigenvalues of global covariance pooling for fine-grained visual recognition,'' \emph{IEEE Trans. Pattern Anal. Mach. Intell.}, vol.~45, no.~3, pp. 3554--3566, 2022.

\bibitem{niu2016ordinal}
Z.~Niu \emph{et~al.}, ``Ordinal regression with multiple output {CNN} for age estimation,'' in \emph{CVPR}, 2016, pp. 4920--4928.

\bibitem{zhu2021convolutional}
H.~Zhu \emph{et~al.}, ``Convolutional ordinal regression forest for image ordinal estimation,'' \emph{{IEEE Trans. Neural Netw. Learn. Syst.}}, vol.~33, no.~8, pp. 4084--4095, 2022.

\bibitem{lei2023core}
Y.~Lei \emph{et~al.}, ``{CORE}: Learning consistent ordinal representations for image ordinal estimation,'' \emph{arXiv: 2301.06122}, 2023.

\bibitem{wang2023controlling}
C.~Wang \emph{et~al.}, ``Controlling class layout for deep ordinal classification via constrained proxies learning,'' in \emph{AAAI}, 2023.

\bibitem{rosati2022novel}
R.~Rosati \emph{et~al.}, ``A novel deep ordinal classification approach for aesthetic quality control classification,'' \emph{Neural Comput. Appl.}, vol.~34, no.~14, pp. 11\,625--11\,639, 2022.

\bibitem{lei2022meta}
Y.~Lei, H.~Zhu, J.~Zhang, and H.~Shan, ``Meta ordinal regression forest for medical image classification with ordinal labels,'' \emph{IEEE-CAA J. Automatica Sin.}, vol.~9, no.~7, pp. 1233--1247, 2022.

\bibitem{lei2023clip}
Y.~Lei \emph{et~al.}, ``{CLIP-Lung}: Textual knowledge-guided lung nodule malignancy prediction,'' in \emph{MICCAI}, 2023.

\bibitem{fonteijn2012event}
H.~M. Fonteijn \emph{et~al.}, ``An event-based model for disease progression and its application in familial {Alzheimer's disease} and {Huntington's} disease,'' \emph{NeuroImage}, vol.~60, no.~3, pp. 1880--1889, 2012.

\bibitem{neyman1934two}
J.~Neyman, ``On the two different aspects of the representative method: The method of stratified sampling and the method of purposive selection,'' \emph{J. R. Stat. Soc.}, vol.~97, no.~4, pp. 558--625, 1934.

\bibitem{cochran1977sampling}
W.~G. Cochran, \emph{Sampling Techniques}, 3rd~ed.\hskip 1em plus 0.5em minus 0.4em\relax John Wiley, 1977.

\bibitem{gong2022ranksim}
Y.~Gong, G.~Mori, and F.~Tung, ``{RankSim}: Ranking similarity regularization for deep imbalanced regression,'' in \emph{ICML}, 2022, pp. 7634--7649.

\bibitem{rolinek2020optimizing}
M.~Rol{\'\i}nek \emph{et~al.}, ``Optimizing rank-based metrics with blackbox differentiation,'' in \emph{CVPR}, 2020, pp. 7620--7630.

\bibitem{poganvcic2020differentiation}
M.~V. Pogan{\v{c}}i{\'c}, A.~Paulus, V.~Musil, G.~Martius, and M.~Rolinek, ``Differentiation of blackbox combinatorial solvers,'' in \emph{ICLR}, 2020.

\bibitem{petersen2010alzheimer}
R.~C. Petersen \emph{et~al.}, ``{Alzheimer's Disease Neuroimaging Initiative (ADNI)}: Clinical characterization,'' \emph{Neurology}, vol.~74, no.~3, pp. 201--209, 2010.

\bibitem{beekly2004national}
D.~L. Beekly \emph{et~al.}, ``The {National Alzheimer's Coordinating Center} {(NACC)} database: an {Alzheimer} disease database,'' \emph{Alzheimer Dis. Assoc. Dis.}, vol.~18, no.~4, pp. 270--277, 2004.

\bibitem{Jang_2022_CVPR}
J.~Jang and D.~Hwang, ``{M3T}: Three-dimensional medical image classifier using multi-plane and multi-slice transformer,'' in \emph{CVPR}, June 2022, pp. 20\,718--20\,729.

\bibitem{fischl2012freesurfer}
B.~Fischl, ``{FreeSurfer},'' \emph{{NeuroImage}}, vol.~62, no.~2, pp. 774--781, 2012.

\bibitem{paszke2019pytorch}
A.~Paszke \emph{et~al.}, ``{PyTorch}: An imperative style, high-performance deep learning library,'' in \emph{NeurIPS}, 2019.

\bibitem{he2015delving}
K.~He, X.~Zhang, S.~Ren, and J.~Sun, ``Delving deep into rectifiers: Surpassing human-level performance on {ImageNet} classification,'' in \emph{CVPR}, 2015, pp. 1026--1034.

\bibitem{song2018collaborative}
G.~Song and W.~Chai, ``Collaborative learning for deep neural networks,'' in \emph{NeurIPS}, 2018.

\bibitem{qian2021my}
S.~Qian \emph{et~al.}, ``Are my deep learning systems fair? an empirical study of fixed-seed training,'' in \emph{NeurIPS}, 2021, pp. 30\,211--30\,227.

\bibitem{korolev2017residual}
S.~Korolev \emph{et~al.}, ``Residual and plain convolutional neural networks for {3D} brain {MRI} classification,'' in \emph{{ISBI}}, 2017, pp. 835--838.

\bibitem{carreira2017quo}
J.~Carreira and A.~Zisserman, ``Quo vadis, action recognition? a new model and the {Kinetics} dataset,'' in \emph{{CVPR}}, July 2017, pp. 6299--6308.

\bibitem{zhang2023video4mri}
Y.~Zhang \emph{et~al.}, ``{Video4MRI}: An empirical study on brain magnetic resonance image analytics with {CNN-based} video classification frameworks,'' \emph{arXiv: 2302.12688}, 2023.

\bibitem{hu2018squeeze}
J.~Hu, L.~Shen, and G.~Sun, ``Squeeze-and-excitation networks,'' in \emph{CVPR}, 2018, pp. 7132--7141.

\bibitem{zhou2020rethinking}
H.~Zhou \emph{et~al.}, ``Rethinking soft labels for knowledge distillation: A bias--variance tradeoff perspective,'' in \emph{ICLR}, 2021.

\bibitem{li2022asymmetric}
X.-C. Li \emph{et~al.}, ``Asymmetric temperature scaling makes larger networks teach well again,'' in \emph{NeuIPS}, 2022, pp. 3830--3842.

\bibitem{qiao2022ranking}
H.~Qiao \emph{et~al.}, ``Ranking convolutional neural network for {Alzheimer’s} disease mini-mental state examination prediction at multiple time-points,'' \emph{Comput. Meth. Programs Biomed.}, vol. 213, p. 106503, 2022.

\bibitem{hinton2015distilling}
G.~Hinton, O.~Vinyals, and J.~Dean, ``Distilling the knowledge in a neural network,'' \emph{arXiv:1503.02531}, 2015.

\bibitem{nachar2008mann}
N.~Nachar \emph{et~al.}, ``The mann-whitney u: A test for assessing whether two independent samples come from the same distribution,'' \emph{Tutorials Quant. Methods Psychol.}, vol.~4, no.~1, pp. 13--20, 2008.

\bibitem{selvaraju2017grad}
R.~R. Selvaraju \emph{et~al.}, ``{Grad-CAM}: Visual explanations from deep networks via gradient-based localization,'' in \emph{CVPR}, 2017, pp. 618--626.

\bibitem{rolls2020automated}
E.~T. Rolls, C.-C. Huang, C.-P. Lin, J.~Feng, and M.~Joliot, ``Automated anatomical labelling atlas 3,'' \emph{NeuroImage}, vol. 206, p. 116189, 2020.

\bibitem{fjell2014normal}
A.~M. Fjell \emph{et~al.}, ``What is normal in normal aging? effects of aging, amyloid and {Alzheimer's} disease on the cerebral cortex and the hippocampus,'' \emph{Prog. Neurobiol.}, vol. 117, pp. 20--40, 2014.

\bibitem{convit2000atrophy}
A.~Convit \emph{et~al.}, ``Atrophy of the medial occipitotemporal, inferior, and middle temporal gyri in non-demented elderly predict decline to {Alzheimer’s} disease,'' \emph{Neurobiol. Aging}, vol.~21, no.~1, pp. 19--26, 2000.

\bibitem{chan2001patterns}
D.~Chan \emph{et~al.}, ``Patterns of temporal lobe atrophy in semantic dementia and {Alzheimer's} disease,'' \emph{Ann. Neurol.}, vol.~49, no.~4, pp. 433--442, 2001.

\bibitem{minoshima1997metabolic}
S.~Minoshima \emph{et~al.}, ``Metabolic reduction in the posterior cingulate cortex in very early {Alzheimer's} disease,'' \emph{Ann. Neurol.}, vol.~42, no.~1, pp. 85--94, 1997.

\bibitem{zhang2022improving}
S.~Zhang, L.~Yang, M.~B. Mi, X.~Zheng, and A.~Yao, ``Improving deep regression with ordinal entropy,'' in \emph{ICLR}, 2023.

\bibitem{faivishevsky2008ica}
L.~Faivishevsky and J.~Goldberger, ``{ICA} based on a smooth estimation of the differential entropy,'' in \emph{NeurIPS}, 2008.

\bibitem{van2008visualizing}
L.~Van~der Maaten and G.~Hinton, ``Visualizing data using {$t$-SNE}.'' \emph{J. Mach. Learn. Res.}, vol.~9, no.~11, 2008.

\bibitem{li2021learning}
W.~Li \emph{et~al.}, ``Learning probabilistic ordinal embeddings for uncertainty-aware regression,'' in \emph{CVPR}, 2021, pp. 13\,896--13\,905.

\bibitem{zhu2020ordinal}
H.~Zhu \emph{et~al.}, ``Ordinal distribution regression for gait-based age estimation,'' \emph{Sci. China-Inf. Sci.}, vol.~63, pp. 1--14, 2020.

\bibitem{yu2023retinal}
Z.~Yu \emph{et~al.}, ``Retinal age estimation with temporal fundus images enhanced progressive label distribution learning,'' in \emph{MICCAI}.\hskip 1em plus 0.5em minus 0.4em\relax Springer, 2023, pp. 629--638.

\bibitem{aisen2010clinical}
P.~S. Aisen \emph{et~al.}, ``Clinical core of the {Alzheimer's} disease neuroimaging initiative: progress and plans,'' \emph{Alzheimers. Dement.}, vol.~6, no.~3, pp. 239--246, 2010.

\bibitem{chen2023berdiff}
T.~Chen, C.~Wang, and H.~Shan, ``{BerDiff}: Conditional {Bernoulli} diffusion model for medical image segmentation,'' \emph{MICCAI}, 2023.

\bibitem{schmidt2023probabilistic}
A.~Schmidt, P.~Morales-{\'A}lvarez, and R.~Molina, ``Probabilistic modeling of inter-and intra-observer variability in medical image segmentation,'' in \emph{ICCV}, 2023, pp. 21\,097--21\,106.

\end{thebibliography}
\end{document}